\documentclass[aps,prb,twocolumn,showpacs]{revtex4}
\usepackage{amsmath,amssymb,mathrsfs}
\usepackage{epsfig}

\newcommand{\eqnref}[1]{Eq.~(\ref{#1})}

\newcommand{\figref}[1]{Fig.~\ref{#1}}
\newcommand{\figsref}[1]{Figs.~\ref{#1}}
\newcommand{\Figref}[1]{Figure~\ref{#1}}
\newcommand{\Figsref}[1]{Figures~\ref{#1}}

\newcommand{\secref}[1]{Sec.~\ref{#1}}
\newcommand{\Secref}[1]{Section~\ref{#1}}

\newcommand{\bfl}{\mathbf{l}}
\newcommand{\bfA}{\mathbf{A}}

\newcommand{\varH}{\mathscr{H}}

\newcommand{\im}{\mathrm{Im}}

\newcommand{\avg}[1]{\left\langle#1\right\rangle}

\begin{document}

\title{Quantum Phase Transitions and Persistent Currents in
  Josephson-Junction Ladders}

\author{Minchul Lee}%
\affiliation{Department of Physics, Seoul National University,
  Seoul 151-747, Korea}%
\affiliation{Department of Physics, Korea University,
  Seoul 136-701, Korea}%

\author{Mahn-Soo Choi}%
\thanks{To whom correspondences should be addressed.
  E-mail: choims@korea.ac.kr}%
\affiliation{Department of Physics, Korea University,
  Seoul 136-701, Korea}%

\author{M.Y. Choi}%
\affiliation{Department of Physics, Seoul National University,
  Seoul 151-747, Korea}%
\affiliation{Korea Institute for Advanced Study, Seoul 130-012, Korea}%

\begin{abstract}
In this work we study quantum phase transitions and persistent
currents in capacitively coupled one-dimensional Josephson-junction
arrays.  We will focus particularly on the roles of exciton-like pairs
in the strong coupling limit in the presence of external gate charges
and magnetic fluxes.  We use the numerical density-matrix
renormalization group method for the study in the full range of values
of gate charge and magnetic flux.  To clarify the various effects, we
report the pair correlation functions and the exciton densities as
welll as the persistent current.
\end{abstract}


\maketitle

\section{Introduction}

Systems of Josephson junctions between small superconducting grains
have been attracting considerable interest for more than two decades.
One of the main attractive features is that they exhibit
manifestations of various phenomena in diverse fields of
condensed-matter physics.  A popular example in contemporary
mesoscopic physics is the Coulomb blockade effect and single charge
(electron or Cooper pair) tunneling.\cite{CoulombBlockade,Schon90}
Persistent current, another hot topic in mesoscopic physics, can also
be embodied in Josephson-junction systems. Since the superconducting
coherence is easily maintained over a macroscopic length scale, a
``necklace'' of Josephson junctions (i.e., one-dimensional periodic
array of Josephson junctions) may be a good testbed for persistent
currents.\cite{Choi93} Moreover, charge fluctuations present in such
systems may induce quantum phase
transitions,\cite{1DQPT,Fazio91,Otterlo93,MQPC,KimBJ} providing a
prototype model for the noble many-body phenomena in strongly
correlated electron systems.\cite{Sachdev98} Another important and
appealing feature of the systems is the experimental tunability: They
not only make mesoscopic devices on their own\cite{Sohn97,Schon90} but
also allow us to test and understand otherwise very subtle points of
interacting many-particle systems,\cite{Sachdev98} which is important
from a fundamental point of view.

Here we consider a particular geometry of Josephson-junction systems:
a ladder of two capacitively coupled one-dimensional (1D)
Josephson-junction arrays.  In the Coulomb blockade regime, a single
charge cannot tunnel across the junction since it is energetically
unfavorable.  Transport is therefore dominated by more complex
elementary processes that involve several charge-tunneling events.
For the particular type of coupling through large inter-array
capacitances, the relevant elementary process in the absence of the
gate charge consists of cotunneling of bound pairs of excess and
deficit charges, which we call ``particles''(excess charges) and
``holes'' (deficit charges), respectively.~\cite{endnote1} It was
first demonstrated on capacitively coupled normal-metal tunneling
junctions\cite{Averin91,Matters97} and later on superconducting
junction arrays.\cite{ChoiMS1D98,Shimada00} In the presence of gate
voltage applied between the electrode islands and the substrate, the
particle-hole symmetry is broken and the particle-hole pair no longer
makes the lowest charging-energy configuration.  For example, when the
particle-hole symmetry is broken maximally (corresponding to the gate
charge given by one half of the elementary charge $2e$), the transport
is governed by cotunneling of particle-void pairs (with the
\emph{void} denoting the absence of any excess or deficit Cooper
pair).\cite{ChoiMS1D98} It is noted that these particle-hole pairs or
particle-void pairs are reminiscent of \emph{excitons}, i.e., bound
states of a band electron and a hole, in solids.  In the previous
work,\cite{ChoiMS1D98} quantum phase transitions induced by the
cotunneling of particle-hole pairs and particle-void pairs near the
particle-hole symmetry line and the maximal-frustration line,
respectively, have been studied by means of perturbative methods.
However, properties of the transport or phase transitions in between
have not been studied.

Effects of an external magnetic flux threading the loop of a ladder of
two capacitively coupled Josephson-junction necklaces (CCJJNs) (see
Fig.~\ref{fig(ccjjn):necklaces}) are even more sophisticated since the
objects involved in the persistent current are not single charges.
Unlike most studies of the persistent current (or equivalently, the
underlying Aharonov-Bohm effect), which focus on single-charged
particles, recent researches into a nano-structure with
non-simply-connected geometry have demonstrated\cite{ABofExciton} that
excitons can contribute to persistent currents, in spite of their
charge neutrality.  The nonvanishing persistent currents in the
system, is attributed to the finite probability of breaking and
recombination of an exciton via intermediate single particle/hole
states.  It is thus quite intriguing to investigate persistent
currents in CCJJNs, where cotunneling of the particles and holes of
Cooper pairs dominates the transport phenomena.  Additional advantage
of the CCJJNs is that the particle-hole or particle-void pairs are
stable while the excitons in semiconductor nano-rings usually have a
finite and short life time.  Notice further that the CCJJNs are
already within the reach of experimental realization.\cite{Shimada00}

In this work we study quantum phase transitions and persistent
currents in a ladder of CCJJNs.  We focus particularly on the roles of
``excitons'' in the presence of the charge frustration due to an
external gate voltage and the magnetic frustration due to an external
magnetic flux threading the necklaces.  We use the numerical
density-matrix renormalization group (DMRG) method\cite{DMRG} to probe
the full ranges of the gate charge and the magnetic flux.
Although we are mainly interested in the strong-coupling limit, we
will consider for comparison both the two limiting cases: decoupled
and strongly coupled cases.  In the limit of strong coupling, we
identify two different superfluid phases, characterized by
condensation of either particle-hole pairs or particle-void pairs,
depending on the gate charge.  In order to disclose properties of the
superfluid phases and formation of excitons explicitly, we measure the
pair correlation function and the exciton density.  The behavior of
the persistent current calculated for small systems reveal the
transport via the separation and recombination process for small
Josephson energies.  At larger Josephson energies, however, the
transport is governed by mixing of low-lying charge states with
higher-energy states.  Finally, we propose an experimentally
realizable system to demonstrate the cotunneling process of excitons.
The intermediate coupling regime is also interesting and more feasible
experimentally.\cite{Shimada00} Unfortunately, however, the numerical
DMRG study in this case is beyond the current computing power,
requiring far more memory than available.  We thus leave the
intermediate region for the future study.

The remaining part of this paper is organized as follows: In
\secref{sec(ccjjn):model} we first describe the model Hamiltonians and
discuss qualitatively the relevant low-energy charge states.  Quantum
phase transitions in a single Josephson-junction necklace and
capacitively coupled Josephson-junction necklaces are examined in
\secref{sec(ccjjn):qpt}.  \Secref{sec(ccjjn):pc} is devoted to the
investigation of the persistent currents in the system, revealing the
AB effect of excitons.  Finally, we summarize the main results in
\secref{sec(ccjjn):conclusion}.

\section{Model\label{sec(ccjjn):model}}

\begin{figure}
\centering%
\epsfig{file=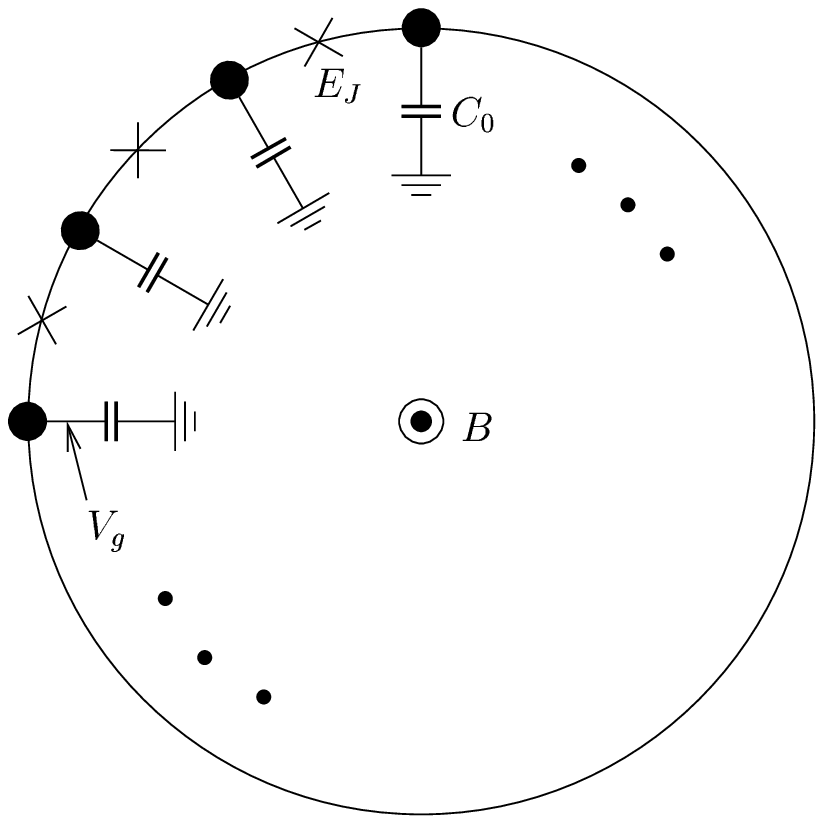,width=5cm}\\ (a)\\[1cm]
\epsfig{file=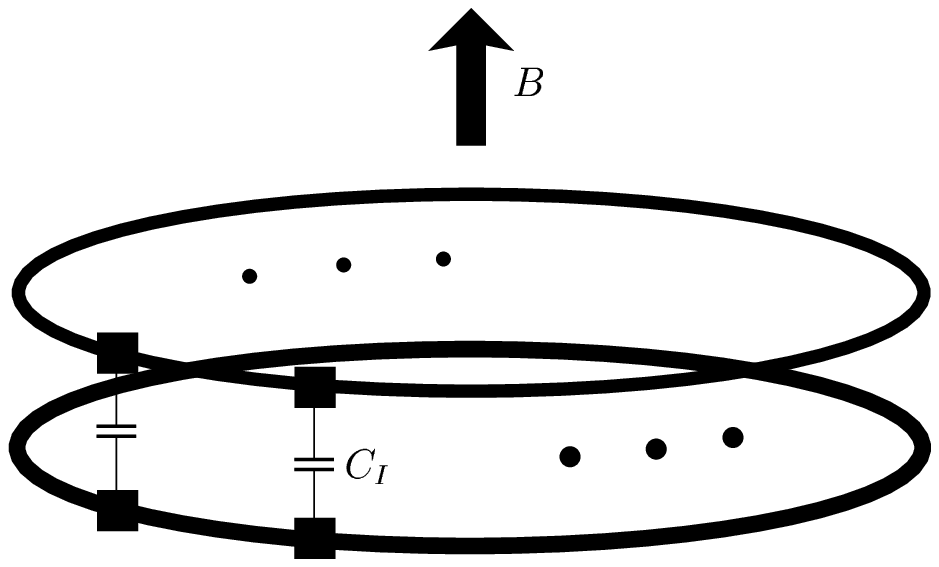,width=5cm}\\ (b)
\caption{Schematic diagrams of (a) a single Josephson-junction
  necklace (from above) and (b) two capacitively coupled necklaces
  (from the diagonal direction).  In (b) each thick ring represents
  the single necklace depicted in (a).}
  \label{fig(ccjjn):necklaces}
\end{figure}

We consider two 1D periodic arrays, which we call {\em necklaces}, of
$N$ superconducting grains as shown in \figref{fig(ccjjn):necklaces}.
Any two nearest-neighboring grains on one necklace form a Josephson
junction of coupling strength $E_J$.  The two necklaces are coupled
with each other via capacitance $C_I$ between corresponding grains, to
form a ``ladder''.  Uniform gate voltage $V_g$ is applied to each
grain through its self-capacitance $C_0$, inducing gate charge $Q =
C_0 V_g$ on each grain.  For convenience, we measure the charge in
units of $2e$, and write $Q\equiv 2en_g$.  In addition, a transverse
magnetic flux threads each necklace.  Such a system is described by
the Hamiltonian
\begin{multline}
\label{eq(ccjjn):Hqpm}
\varH = 2e^2 \sum_{lx,l'x'} [n^l_x - n_g]
C^{-1}_{lx,l'x'} [n^l_{x'} - n_g] \\
- E_J \sum_{lx} \cos(\phi^l_x - \phi^l_{x{+}1} - A_x) \,,
\end{multline}
where the number $n^l_x$ of the Cooper pairs and the phase $\phi^l_x$
of the superconducting order parameter at site $x$ on the $l$th
necklace ($l=1,2$) are quantum-mechanically conjugate variables:
\begin{math}
[n^l_x,\phi^{l'}_{x'}] = i \delta_{l,l'}\delta_{x,x'}
\end{math}.
The bond angle $A_x$ is given by the line integral of the vector
potential $\bfA$ introduced by the applied magnetic field:
\begin{equation}
A_x = \frac{2\pi}{\Phi_0} \int_x^{x+1} d\bfl\cdot\bfA = \frac{2\pi f}{N},
\end{equation}
where $f$ denotes the total flux in units of the flux quantum $\Phi_0
\equiv 2\pi\hbar c/2e$.  Assuming that junction capacitances are
negligible, we write the capacitance matrix $C_{lx,l'x'}$ in the
form~\cite{Simanek94a}
\begin{equation}
C_{lx,l'x'} = \left[
  C_0 \delta_{l,l'} + C_I(2\delta_{l,l'}-1)
\right] \delta_{x,x'}
\equiv C_{ll'} \delta_{x,x'}
\end{equation}
and also define charging energy scales $E_0 \equiv e^2/2C_0$ and
$E_I\equiv e^2/2C_I$, associated with the corresponding capacitances.
Notice that when $n_g=0$ the Hamiltonian in Eq.~\eqref{eq(ccjjn):Hqpm}
is symmetric with respect to the particle-like (excess Cooper pairs)
excitations and hole-like (deficit pairs) ones.  On the other hand,
charges on each grain are maximally frustrated when $n_g=1/2$.  For
later use, we thus name the lines corresponding to $n_g=0$ and
$n_g=1/2$ the particle-hole symmetry line and the maximal-frustration
line, respectively, in the phase diagram.

For the DMRG analysis, we represent the Hamiltonian in
\eqnref{eq(ccjjn):Hqpm} in the boson number basis.  Based on the
commutation relation between the number $n^l_x$ and the phase
$\phi^l_x$, we identify the boson creation operator $b^{l\dag}_x$ at
site $x$ on necklace $l$ with $e^{i\phi^l_x}$.  In terms of the boson
operators, we thus obtain the Bose-Hubbard Hamiltonian
\begin{multline}
\label{eq(ccjjn):H}
\varH_{BH} = \frac{8E_0}{2} \sum_{ll'x} n^l_x C_0C_{ll'}^{-1} n^{l'}_x
- 8E_0(n_g+\bar{n})\sum_{lx} n^l_x \\
- \frac{E_J}{2\bar{n}} \sum_{lx} \left( e^{-2\pi i f/N} b^{l\dag}_x
  b^l_{x{+}1} + h.c\right),
\end{multline}
where $\bar{n}$ is the average boson number per site.  Note that in
the quantum phase model the Josephson energy term is independent of
the number fluctuations, while the corresponding (hopping) term is not
in the Bose-Hubbard model.  To alleviate the effects of number
fluctuations in the Bose-Hubbard model, we consider the case that the
average boson number $\bar{n}$ per site is large\cite{QPMtoBHM}:
Throughout this study we set $\bar{n}$ to be 10000.

\begin{figure}[!t]
\centering%
\epsfig{file=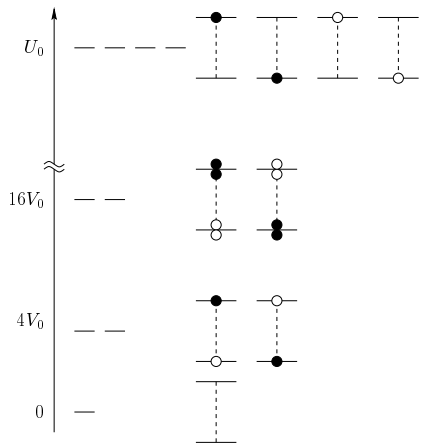,height=6cm}
\caption{Energy levels of the charging energy part in
  \eqnref{eq(ccjjn):Hc} and corresponding charge configurations near
  the particle-hole symmetry line. Filled and empty circles denote
  particles and holes, respectively; paired (upper and lower) solid
  lines represent the two coupled arrays, the couplings between which
  are illustrated by the dashed lines. The low-lying energy levels
  satisfying $n_x^+ = 0$ are well separated by a large amount of
  energy (of the order of $E_0$) from those with $n_x^+ \ne 0$.}
  \label{fig(ccjjn):el_phs}
\end{figure}

\begin{figure}[!t]
\centering%
\epsfig{file=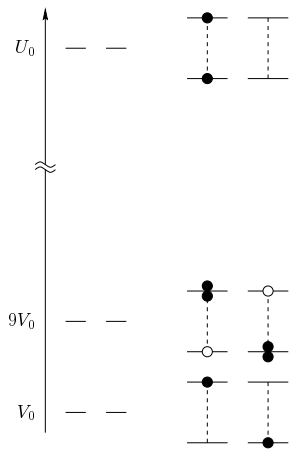,height=6cm}
\caption{Energy levels and corresponding charge configurations near the
  maximal-frustration line.  Note that the ground state is two-fold
  degenerate per site.}
\label{fig(ccjjn):el_mf}
\end{figure}

Capacitive coupling between necklaces drastically affects the
low-lying charge excitations, especially, in the strong coupling
regime.  To examine the charge configurations in the low-lying states,
it is convenient to rewrite the charging energy part in the
Hamiltonian (\ref{eq(ccjjn):H}):
\begin{multline}
\label{eq(ccjjn):Hc}
\varH_{BH} = U_0 \sum_x (n_x^+ - 2n_g)^2 + V_0 \sum_x (n_x^-)^2 \\
- \frac{E_J}{2\bar{n}} \sum_{lx} \left( e^{-2\pi i f/N} b^{l\dag}_x
  b^l_{x{+}1} + h.c\right) \,,
\end{multline}
where we have defined new energy scales $U_0 \equiv 2E_0$ and $V_0
\equiv 2E_0/(1+2C_I/C_0)$ and charge variables
\begin{math}
n_x^\pm \equiv (n_{1,x} - \bar{n}) \pm (n_{2,x} - \bar{n})
\end{math}.
Note that $n_x^+$ represents the total number of (excess) Cooper pairs
on the $x$th rung of the ladder.  In the regime of concern ($C_I\gg
C_0$, i.e., $E_I\ll E_0$), we have $U_0\gg{}V_0$ and $U_0\gg{}E_J$,
and the charge configurations satisfying $n_x^+-2n_g\simeq 0$ are thus
strongly favored.  In such a charge configuration $n_x^-/2$
corresponds to the number of excitons (particle-hole or particle-void
pairs, see below).

The representation in Eq.~\eqref{eq(ccjjn):Hc} of the Hamiltonian
allows us to distinguish clearly the two interesting regions from each
other: near the particle-hole symmetry line ($n_g=0$) and near the
maximal-frustration line ($n_g=1/2$), as one can observe from the
energy spectra of the charging energy part illustrated in
\figsref{fig(ccjjn):el_phs} and \ref{fig(ccjjn):el_mf} for the two
regimes, respectively.  Near the maximal-frustration line, the charge
configurations that do not satisfy the condition $n_x^+=1$ (for all
$x$) have a huge excitation gap of the order of $E_0$.  Furthermore,
the ground states of the charging energy part, separated from the
excited states by a gap of the order of $E_I$, have two-fold
degeneracy for each $x$, corresponding to $n_x^- = \pm1$.  Near the
particle-hole symmetry line, on the other hand, low-energy charge
configurations should satisfy the condition $n_x^+ = 0$ for all $x$.
Unlike the former case, the ground state of the charging energy is
non-degenerate and forms a Mott insulator characterized by $n_{1,x}=
n_{2,x} = 0$ for all $x$.  As $E_J$ is turned on, the ground state is
mixed with the states with $n_x^- = \pm 2$.  In the intermediate
region $(0<n_g<1/2)$, these two kinds of energy spectra are
interleaved to form a complex shape of the energy levels.

\section{Quantum phase transitions\label{sec(ccjjn):qpt}}

The competition between charge order and phase coherence gives rise to
quantum fluctuations and quantum phase transitions at zero
temperature.  For large charging energy ($E_0 \gg E_J$), the bosons
become localized and the system is in the Mott insulator phase with
integer density. On the contrary, for large hopping energy or
Josephson energy ($E_J \gg E_0$), coherence of the phases $\phi_x$
dominates over the system and the superfluid (SF) region with
delocalized bosons is observed.  The properties and universality
classes of the phase transitions, however, depend strongly on the
coupling strength $C_I/C_0$ as well as the chemical potential $\mu
\equiv 8E_0(n_g+\bar{n})$.  The charge frustration $n_g$ may be
restricted to the range $[1,1/2]$ since the Hamiltonian in
\eqnref{eq(ccjjn):Hc} is periodic in $n_g$ with period unity and has
reflection symmetry about the $n_g=1$ (or any integer) line. In the
followings we investigate two limiting cases: the decoupled case ($C_I
= 0$) and the strongly coupled one ($C_I \gg C_0$).

\subsection{Single Josephson-junction necklace\label{sec(ccjjn):qpt_jjn}}

The phase transition in a single Josephson-junction array has been
studied quite extensively and it has been found that its nature
depends crucially on the gate voltage.  In the presence of nonzero
gate voltage ($n_g\ne0$), the density of the system changes as the
phase boundary is crossed from the incompressible insulator to the
compressible superfluid.  The transition can thus be located at the
point where in the thermodynamic limit the density of the ground state
becomes different from one of the insulator ground state as the
Josephson energy is increased. On the other hand, in the particle-hole
symmetry line ($n_g=0$), the density remains to be an integer at the
phase transition.  Therefore, in this case the phase boundary is
determined by the single-particle excitation gap.  This is possible
because in the superfluid phase the ground state is a superposition of
states with different boson numbers, the energy gap between the ground
state and the states with additional particles, which is finite in the
insulating phase, vanishes in the superfluid phase.

Since the Hamiltonian conserves total charge number, the DMRG
algorithm can be set up to target states with given total excess
number $M$ of bosons.  We thus obtain the phase diagram of the system
by comparing energies of the ground states with different boson
numbers: the energy $E_{M=0}$ of the insulator ground state with zero
excess boson density ($\avg{n_x} = \bar{n}$) and the energy $E_{M=1}$
of the eigenstate with an additional particle upon the ground state.
Through the linear extrapolation of the energy gap $E_{M=1}-E_{M=0}$
for finite system size $N=64,128$, and $256$, we have estimated the
gap in the thermodynamic limit and located the phase boundary at the
point where the gap is zero. For high numerical accuracy and access to
large systems, the finite-size DMRG algorithm and open boundary
conditions have been used.  During the DMRG process, the boson number
at each site is truncated to be less than six and the discarded weight
is set to be less than $10^{-6}$, giving rise to negligible errors in
the gap energy.\cite{Kuhner00} The magnetic frustration $f$ is set
equal to zero because it can be gauged away and becomes irrelevant in
the thermodynamic limit.

\begin{figure}[!t]
\centering%
\epsfig{file=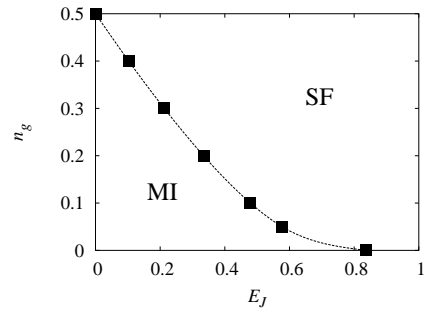,width=6cm}
\caption{Phase diagram for a single 1D Josephson array. The phase
  boundary separates the Mott insulator (MI) with zero excess Cooper
  pair density from the superfluid phase (SF).  The Josephson energy
  $E_J$ is expressed in units of $8E_0$ and the error bars in the
  $E_J$ direction (for given $n_g$) are smaller than the square
  symbols.  The line is merely a guide to the eye.}
\label{fig(ccjjn):jjnpd}
\end{figure}

\Figref{fig(ccjjn):jjnpd} displays the resulting phase diagram for the
quantum phase model on the $E_J$-$n_g$ plane, in the range $0\le
n_g\le1/2$ and $0\le E_J \le 8E_0$.  For convenience, here and in all
subsequent figures, the energy is expressed in units of $8E_0$.
The phase diagram, where the Mott insulator region with zero excess
boson density is separated from the compressible superfluid, is in
good agreement with those obtained via the perturbative
expansion\cite{QPMexpansion} and the quantum Monte Carlo
method.\cite{Baltin97} On the particle-hole symmetry line the quantum
phase model is mapped exactly to the (1+1)-dimensional $XY$ model,
predicting a Beresenskii-Kousterlitz-Thouless (BKT)
transition\cite{BKT} driven purely by phase fluctuations.  The
sharp-pointed shape of the insulating region near the symmetry line
reflects the slowness in closing the energy gap in the BKT
transition.\cite{Kuhner00} In case that the particle-hole symmetry is
broken (away from the symmetry line), no such slowness is found and
the commensurate-incommensurate transition belongs to a universality
class other than that of the $XY$ model, with different critical
exponents\cite{Baltin97} and RG characteristics.\cite{Choi01}

\subsection{Strongly coupled Josephson-junction necklaces\label{sec(ccjjn):ptccjjn}}

In the strong coupling limit, the low-energy charging states relevant
to the phase transition are the the particle-hole pairs (with $n_x^+ =
0$ and $n_x^- = \pm2$) and the particle-void pairs (with $n_x^+ = 1$
and $n_x^- = \pm1$).  For small hopping strength, these excitons are
localized and the system is in the Mott insulator phase. As $E_J$
increases, the phase boundary is crossed from the insulator to the
superfluid which, in this case, originates from condensation of the
excitons.  Accordingly, as in the case of a single Josephson-junction
necklace, the transition can be located as one tracks the energy taken
to add an exciton to the insulator: At the phase boundary this energy
vanishes in the thermodynamic limit. Which kind of exciton between the
particle-hole pair and the particle-void pair is relevant depends on
the charge frustration $n_g$. Near the particle-hole symmetry line
($n_g \approx 0$), the particle-hole pairs are energetically favorable
and governs the phase transition.  As $n_g$ is increased and in the
presence of the Josephson tunneling, in contrast, the particle-void
pairs begin to be dominant faster than the particle-hole pairs, which
will be shown below.

In the DMRG procedure we have associated the target state with a pair
of total excess boson numbers $(M_1,M_2)$ on the two arrays by
utilizing the boson number conservation.  In order to locate the phase
boundary, we have calculated the energy $E_{(M_1, M_2)}$ of three
kinds of eigenstate: the insulator ground state with $(M_1,M_2) =
(0,0)$ and the states with additional particle-hole and particle-void
pairs upon the ground state, labeled by $(M_1,M_2)=(1,-1)$ and
$(1,0)$, respectively. We have extrapolated the energy gaps
$E_{(1,-1)} - E_{(0,0)}$ and $E_{(1,0)} - E_{(0,0)}$ for finite system
size $N = $16, 32, and 64 to locate the transition points where the
gaps vanish in the thermodynamic limit. As in the case of a single
Josephson necklace, we have employed the finite-size DMRG algorithm,
imposing open boundary conditions.

\begin{figure}[!t]
\centering%
\epsfig{file=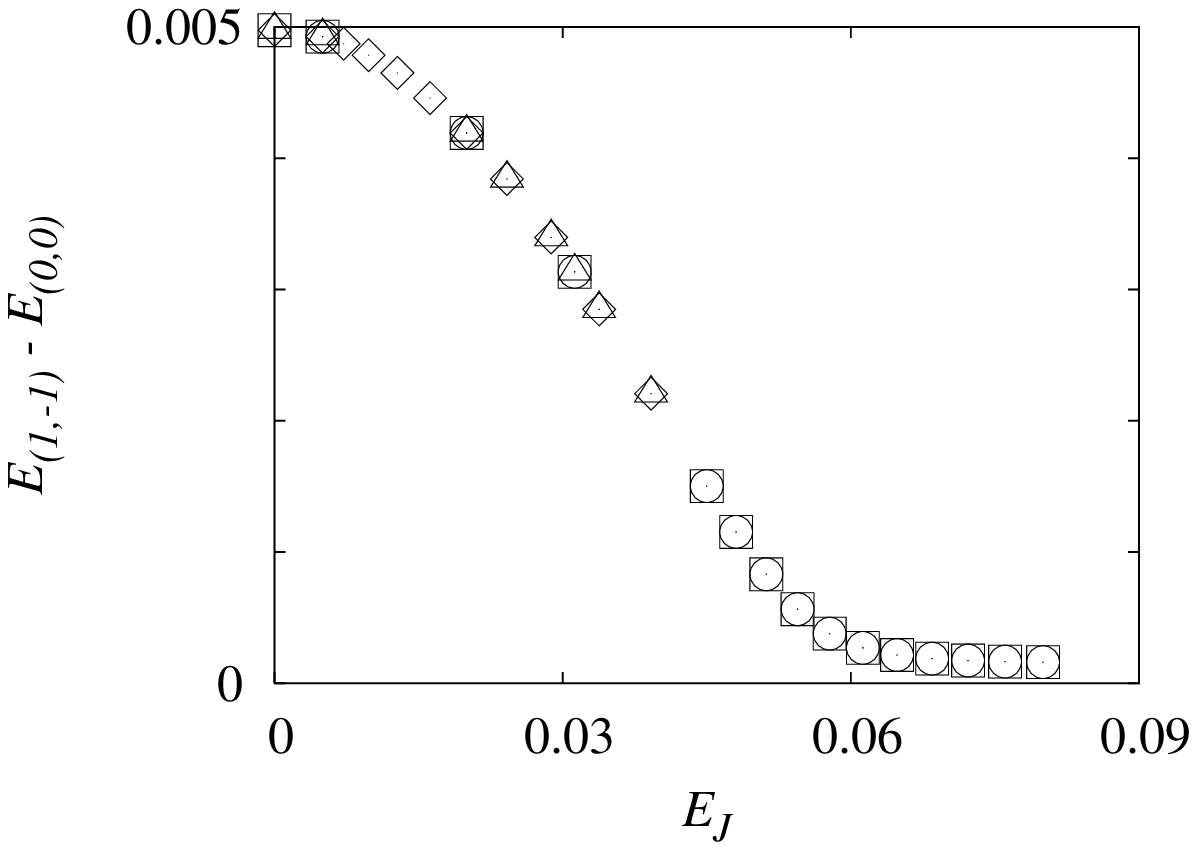,width=6cm}\\ (a) \\
\epsfig{file=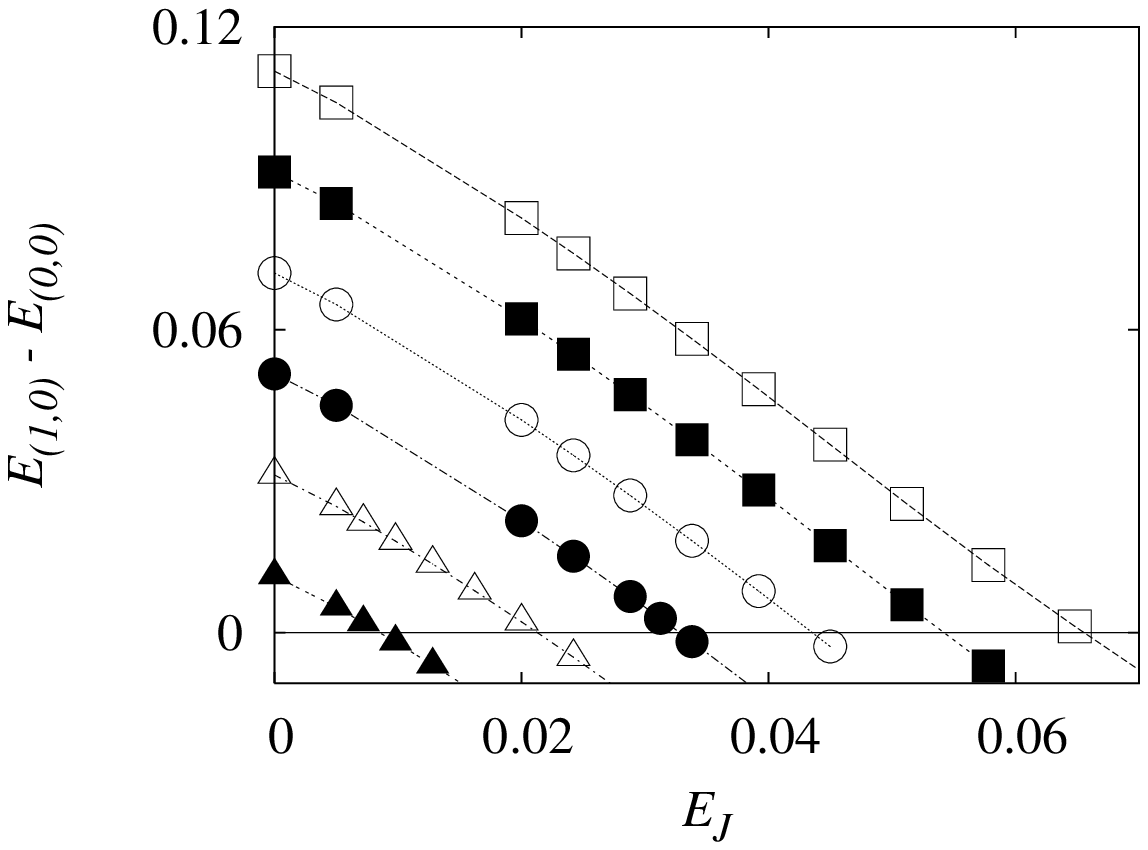,width=6cm}\\ (b)
\caption{Energy gaps (a) $E_{(1,-1)} - E_{(0,0)}$ and (b)
  $E_{(1,0)} - E_{(0,0)}$.  Both are taken for $C_I/C_0 = 100$ at the
  system size $N = 32$.  Each symbol corresponds to a different value
  of charge frustration: (a) $n_g = 0(\square)$, $0.1(\circ)$,
  $0.2(\triangle)$, and $0.24(\lozenge)$; (b) $n_g = 0.14(\square)$,
  $0.16(\blacksquare)$, $0.18(\circ)$, $0.2(\bullet)$,
  $0.22(\triangle)$, and $0.24(\blacktriangle)$.  The energies on both
  axes are expressed in units of $8E_0$.}
\label{fig(ccjjn):es}
\end{figure}

\Figref{fig(ccjjn):es} shows the energy gaps as functions of the
Josephson energy at various charge frustrations, in the system with
$C_I/C_0 = 100$ and $N=32$.  From \figref{fig(ccjjn):es}(a) we observe
that the excitation energy $E_{(1,-1)} - E_{(0,0)}$ for different
charge frustrations collapses into one curve, which also happens at
other system sizes.  This indicates that the critical Josephson energy
at the transition driven by the particle-hole pairs does not depend on
$n_g$.  On the other hand, the energy gap $E_{(1,0)} - E_{(0,0)}$
decreases almost linearly with the increase of $n_g$ and $E_J$, as
shown in \figref{fig(ccjjn):es}(b).  The larger $n_g$ is, the smaller
the Josephson energy $E_J$ at which the energy gap vanishes becomes.
For $n_g\gtrsim0.14$, the critical value of $E_J$ become even less
than that for the particle-hole pairs.

\begin{figure}[!t]
\centering%
\epsfig{file=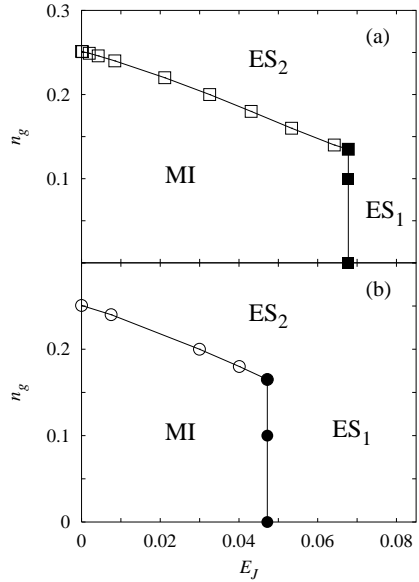,width=6cm}
\caption{Phase diagrams for strongly coupled 1D Josephson arrays
  for (a) $C_I/C_0 = 100$ and (b) $C_I/C_0 = 200$.  Displayed are the
  phase boundaries between the Mott insulator (MI) phase and the
  superfluid phase (ES$_1$ and ES$_2$).  Regions ES$_1$ and ES$_2$ in
  the superfluid phase are distinguished by the dominant transport
  mechanism (see the text).  The transitions across the boundaries
  indicated by filled and empty symbols are driven by the
  particle-hole pairs and by the particle-void pairs, respectively.
  The lines are merely guides to eyes.}
\label{fig(ccjjn):ccjjnpd}
\end{figure}

The resulting phase diagrams for strongly coupled arrays with $C_I/C_0
= 100$ and $200$ are exhibited in \figref{fig(ccjjn):ccjjnpd} (a) and
(b), respectively.  Based on the dominant transport mechanism, one can
distinguish three regions in the superfluid phase: ES$_1$, ES$_2$, and
SFUB.  In region ES$_1$ the transport is driven mainly by the excitons
of particle-hole pairs; in ES$_2$ it is driven by particle-void pairs.
In region SFUB, on the other hand, single-particle processes dominate
the transport in the system.  Such superfluid of unpaired bosons
(SFUB) is to be observed at $E_J/8E_0 \sim 1$, far to the right from
regions ES$_1$ and ES$_2$, and not shown in the phase diagram given by
Fig.~\ref{fig(ccjjn):ccjjnpd}.
We note that different transport mechanisms take over dominant roles
gradually as the control parameters are changed.  Therefore, regions
ES$_1$ and ES$_2$ in Fig.~\ref{fig(ccjjn):ccjjnpd} should not
correspond to truly distinct phases.

Previous studies on mapping of the system at the particle-hold
symmetry line to (1+1)-dimensional system of classical
vortices\cite{ChoiMS1D00,Choi02} insisted that the system is
effectively described by a two-dimensional $XY$ model and exhibits a
BKT transition at the critical Josephson energy $E_J/8E_0|_c =
4K_{BKT}^2 (1+\sqrt{1+2C_I/C_0})^{-2}\approx 2K_{BKT}^2(C_0/C_I)$,
where $K_{BKT} \approx 0.748$ is the critical coupling strength for
the standard $XY$ model.  Our data, though being unable to discern
nature of the transition, show that the critical Josephson energy is
inversely proportional to $\sqrt{C_I/C_0}$ instead of $C_I/C_0$,
apparently favoring against the BKT transition.  This result is quite
reasonable in view of the fact that the cotunneling process of
particle-hole pairs via an intermediate virtual state happens with the
probability proportional to $E_J^2/E_0E_I$, leading to $E_J/E_0|_c
\propto \sqrt{C_I/C_0}$.  In addition, the nonzero charge frustration
does not change the properties of the phase transition abruptly, in
contrast to the case of a single array.  Instead, the transition point
as well as the qualitative properties is preserved up to $n_g \approx
0.135$ for $C_I/C_0 = 100$ and to $n_g \approx 0.165$ for $C_I/C_0 =
200$; there is no increase in the critical value of $E_J$ as predicted
in Ref.~\onlinecite{ChoiMS1D98}.
Since our model neglects the junction capacitance on each necklace,
for
\begin{equation}
n_g > n_g^* \equiv \frac14\left(1+\frac{1}{1+2C_I/C_0}\right),
\end{equation}
each site has two-fold degenerate ground states ($n_x^+ = 1$ and
$n_x^- = \pm1$) of the charging energy.  Accordingly, the Josephson
energy of any strength brings about charge fluctuations to drive the
system into the superfluid phase.  Indeed \Figref{fig(ccjjn):ccjjnpd}
shows that the MI phase ceases to exist for $n_g > n_g^*$, regardless
of $E_J$.  With nonzero junction capacitance, the degeneracy is
expected to be broken, generating another insulating phase: the
charge-density wave (CDW) phase.  The perturbative
study\cite{ChoiMS1D98} has found that as the Josephson energy is
increased the system goes from the CDW insulator to the Luttinger
liquid phase.

\begin{figure}[!t]
\centering%
\epsfig{file=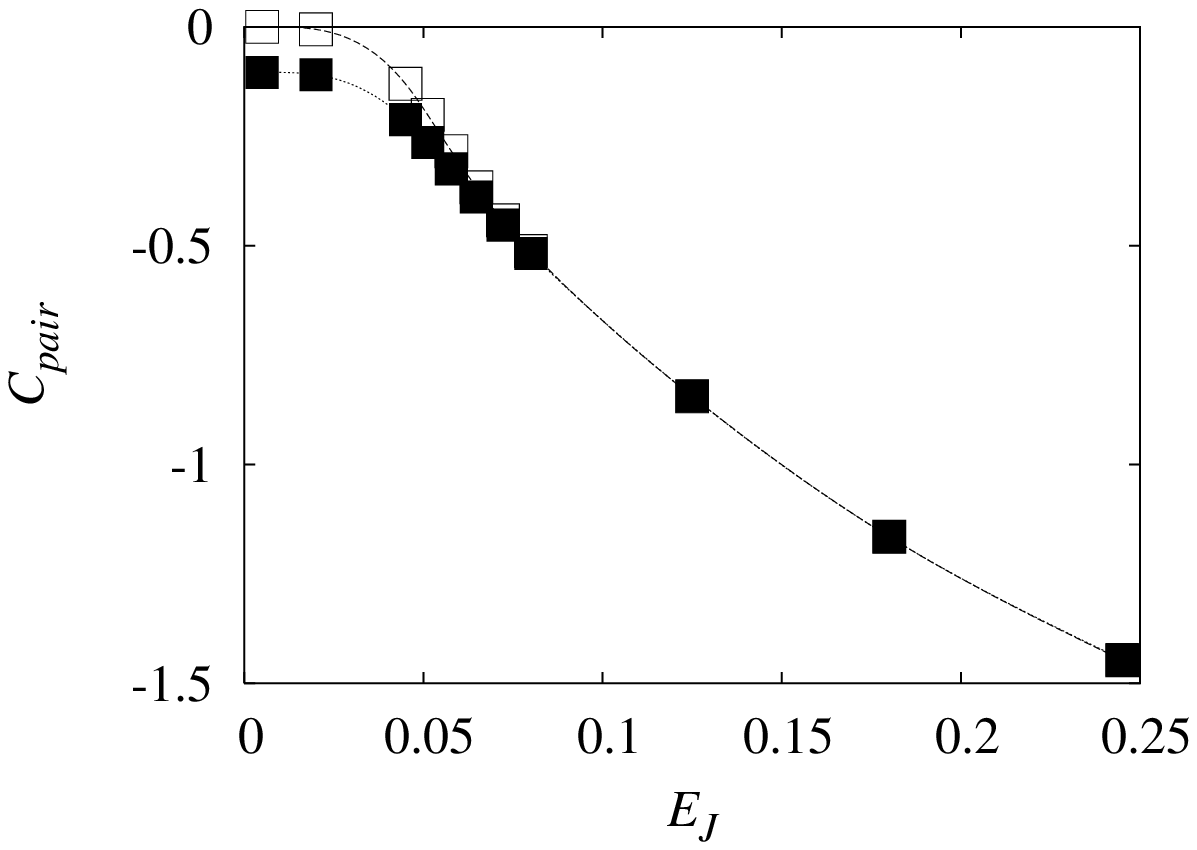,width=6cm}\\ (a) \\
\epsfig{file=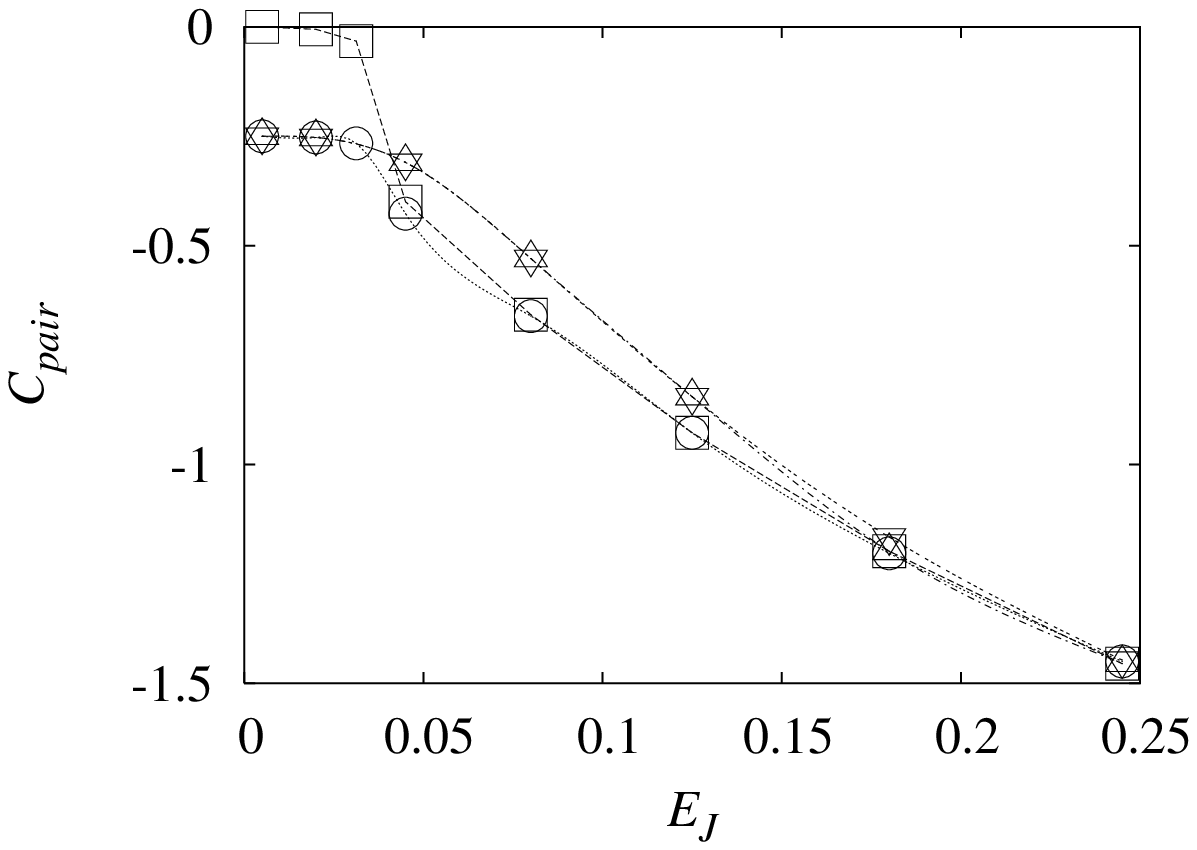,width=6cm}\\ (b)
\caption{Pair correlation $C_{pair}$ versus the Josephson energy.
  (a) For $n_g = 0$, the pair correlation functions in the insulator
  ground state (empty symbols) and in the state with one particle-hole
  pair added upon the ground state (filled symbols) are plotted.  (b)
  shows the correlations in the ground state in the presence of the
  gate voltage $n_g = 0.2(\square), 0.3(\circ), 0.4(\triangle)$, and
  $0.5(\triangledown)$.  Here we set $C_I/C_0 = 100$ and the system
  size $N = 8$, and lines are guides to eyes.}
\label{fig(ccjjn):Cpair}
\end{figure}

\begin{figure}[!t]
\centering%
\epsfig{file=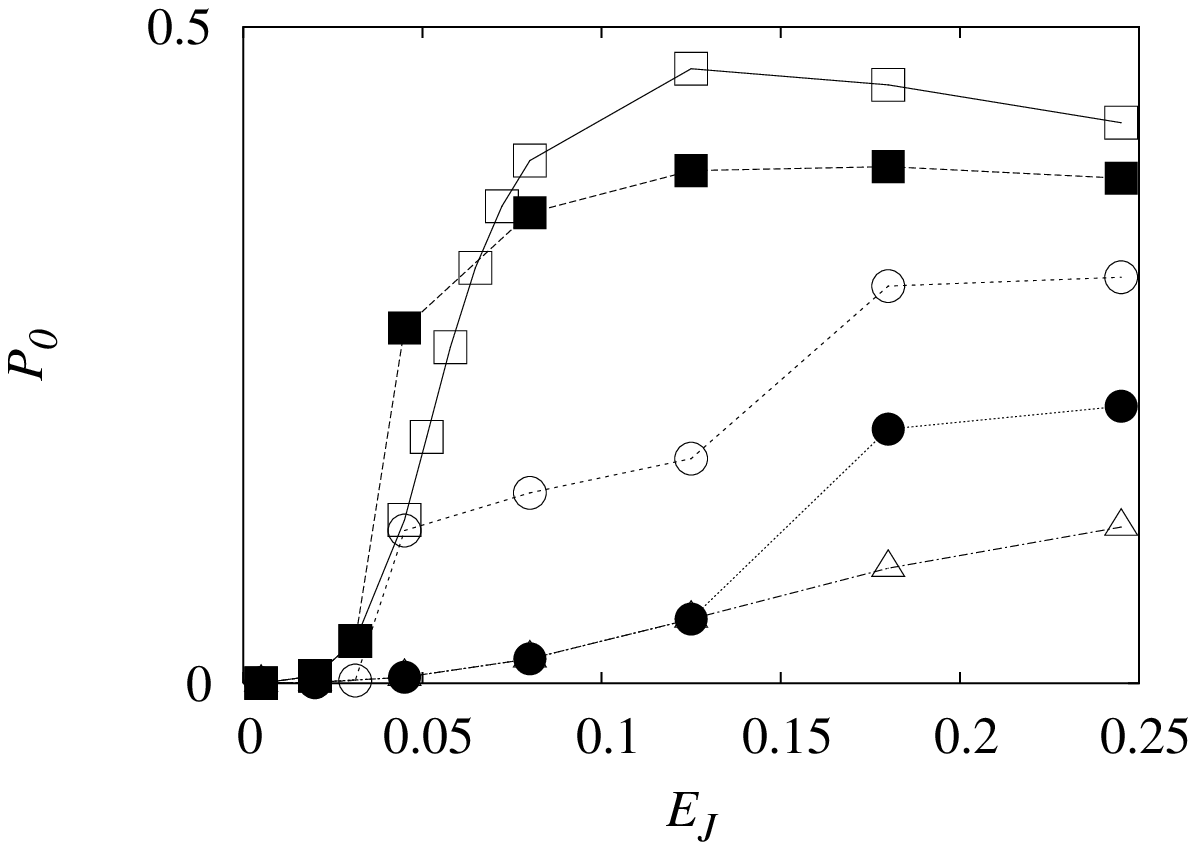,width=6cm}\\ (a) \\
\epsfig{file=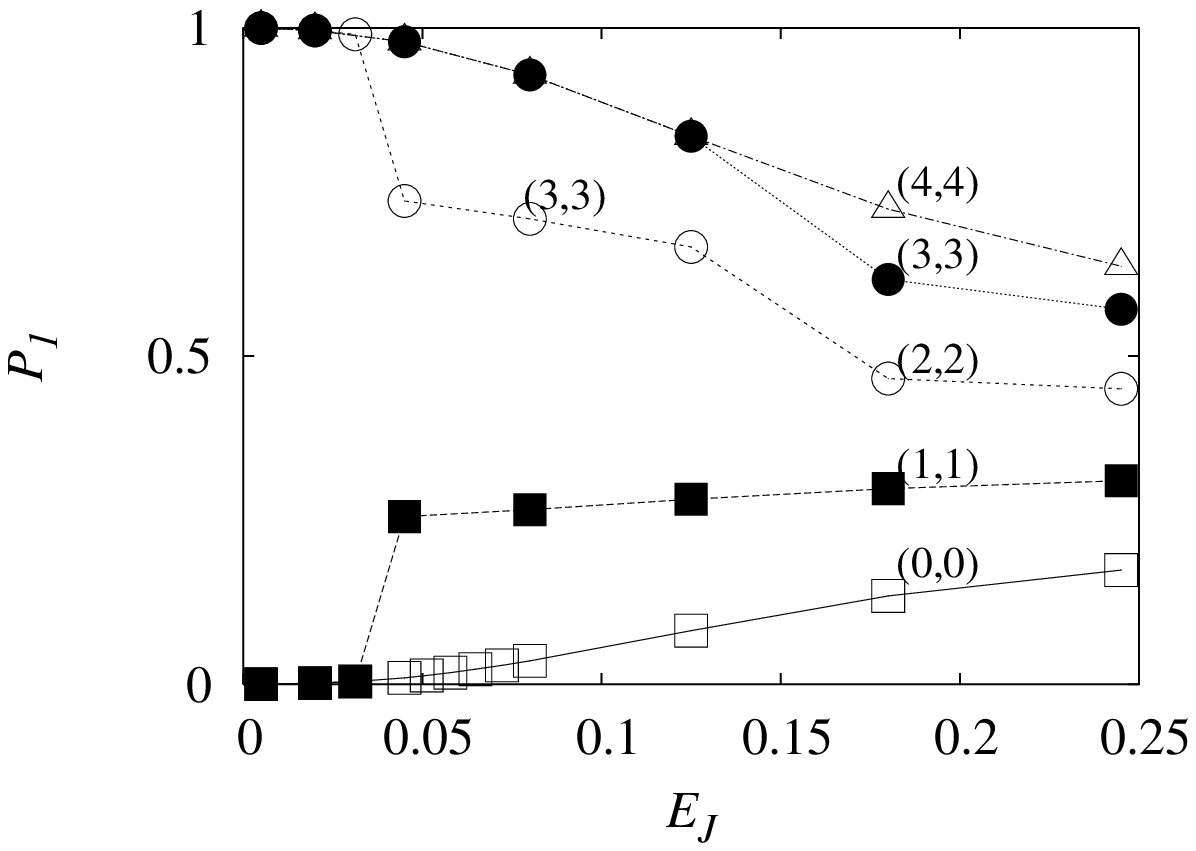,width=6cm}\\ (b)
\caption{Exciton densities (a) $P_0$ and (b) $P_1$ versus the
  Josephson energy for gate voltage $n_g = 0(\square),
  0.2(\blacksquare), 0.3(\circ), 0.4(\bullet)$, and $0.5(\triangle)$.
  Other parameters are the same as those in \figref{fig(ccjjn):Cpair}.
  The pairs of numbers $(M_1,M_2)$ represent the total excess boson
  numbers of the ground state at given parameters.}
\label{fig(ccjjn):P}
\end{figure}

To witness the activity of the excitons in the phase transition, we
have measured the pair correlation function defined to be
\begin{equation}
  C_{pair} \equiv \frac{1}{N} \sum_x \avg{(n_x^1 - \avg{n_x^1})(n_x^2 - \avg{n_x^2})}
\end{equation}
together with the exciton density $P_0$ of the particle-hole pairs and
$P_{\pm1}$ of the particle-void pairs:
\begin{eqnarray}
  P_0 & \equiv &
  \frac{1}{N} \sum_x \avg{\delta_{n_x^1 + n_x^2,0} - \delta_{n_x^1,0}\delta_{n_x^2,0}} \\
  P_{n^+} & \equiv &
  \frac{1}{N} \sum_x \avg{\delta_{n_x^1 + n_x^2,n^+}} \qquad(n^+ \ne 0).
\end{eqnarray}
The pair correlation function assumes zero if there is no correlation
between the boson numbers on the two arrays; a particle-hole or
particle-void pair at every site on the average contributes to
$C_{pair}$ by $-1$ or $-1/4$.

\Figref{fig(ccjjn):Cpair} shows that the pair correlation is negative
and monotonically decreases with $E_J$, which indicates that larger
hopping strength makes more excitons come into the system.  For $n_g =
0$ [see \figref{fig(ccjjn):Cpair}(a)], the two pair correlation
functions, one for the insulator ground state and the other for the
state with an additional particle-hole pair, approach each other and
collapse at $E_J/8E_0 \gtrsim 0.07$, giving another evidence for the
condensation of excitons.  For $n_g = 0.2$, the correlation changes
abruptly at the phase transition, as shown in
\figref{fig(ccjjn):Cpair}(b).  For $n_g \ge 0.3$, we have $C_{pair}
\approx -1/4$ for small values of $E_J$, which reflects the
contribution of particle-void pairs.  Although the correlations for
$n_g \ge 0.3$ become similar at large values of $E_J$, they do not
coincide at intermediate values, implying the difference in the
history of pair generation.

The behavior of the exciton density with $E_J$, displayed in
\figref{fig(ccjjn):P}, manifests more clearly the formation of the
excitons.  The particle-hole pair density $P_0$ grows with $E_J$ [see
\figref{fig(ccjjn):P}(a)], except for the case of $n_g = 0$ and large
$E_J$.  On the other hand, \figref{fig(ccjjn):P}(b) shows that the
density $P_1$ of the particle-void pairs with $n_x^1+n_x^2=1$ has two
kinds of tendencies: For $n_g < n_g^*$ the density $P_1$ increases
with $E_J$; otherwise it decreases.  The steep changes in both
densities $P_0$ and $P_1$ happens when the total excess boson number
$M = M_1+M_2$ of the ground state is altered.  Such behaviors of $P_0$
and $P_1$ reveal that more than one kinds of excitons proliferate in
the system as $E_J$ is increased beyond its critical value.  With
large $E_J$, the kinetic energy gain due to the Josephson tunneling
term compensates for the charging energy gap between different kinds
of excitons.

We close this section with a comment about the pair correlation in the
limit $E_J/8E_0 \gg 1$, which is beyond our current computational
power. Our data shows no indication for the decrease of the pair
correlation with $E_J$ raised.  However, when the Josephson energy is
large enough for the single processes of unpaired particles to
prevail, the pair correlation may eventually approach zero again.

\section{Persistent current\label{sec(ccjjn):pc}}

In this section we consider the persistent current along the
necklaces, induced by the threading external magnetic field. Since
tunneling of Cooper pairs between necklaces is not permitted, the
persistent current carried by each necklace is given by the derivative
of the energy with respect to the magnetic flux $f$:\cite{Choi93}
\begin{equation}
I_l
= \frac{e}{2\pi\hbar}
\left. \avg{\frac{\partial\varH}{\partial f}} \right|_l
= -\frac{eE_J}{N\hbar \bar{n}}
\im\avg{e^{-2\pi i f/N} b^{l\dag}_x b^l_{x{+}1}} \,,
\end{equation}
which is simply the supercurrent through the Josephson junctions. The
current in the system is thus given by the imaginary part of
$\avg{\exp\left[i(\phi^l_x-\phi^l_{x{+}1}-A_x)\right]}$, the real part
of which describes the gauge-invariant phase correlation function
between nearest neighboring grains. Since the current is periodic in
$f$ with period unity and an odd function of the flux $f$, it is
sufficient to calculate the current in the range $f \in [0,1/2]$. As
in the previous section, we focus on two extreme cases: the decoupled
case ($C_I = 0$) and strongly coupled one ($C_I/C_0\gg 1$).

\subsection{Single Josephson-junction necklace}

\begin{figure}[!t]
\centering%
\epsfig{file=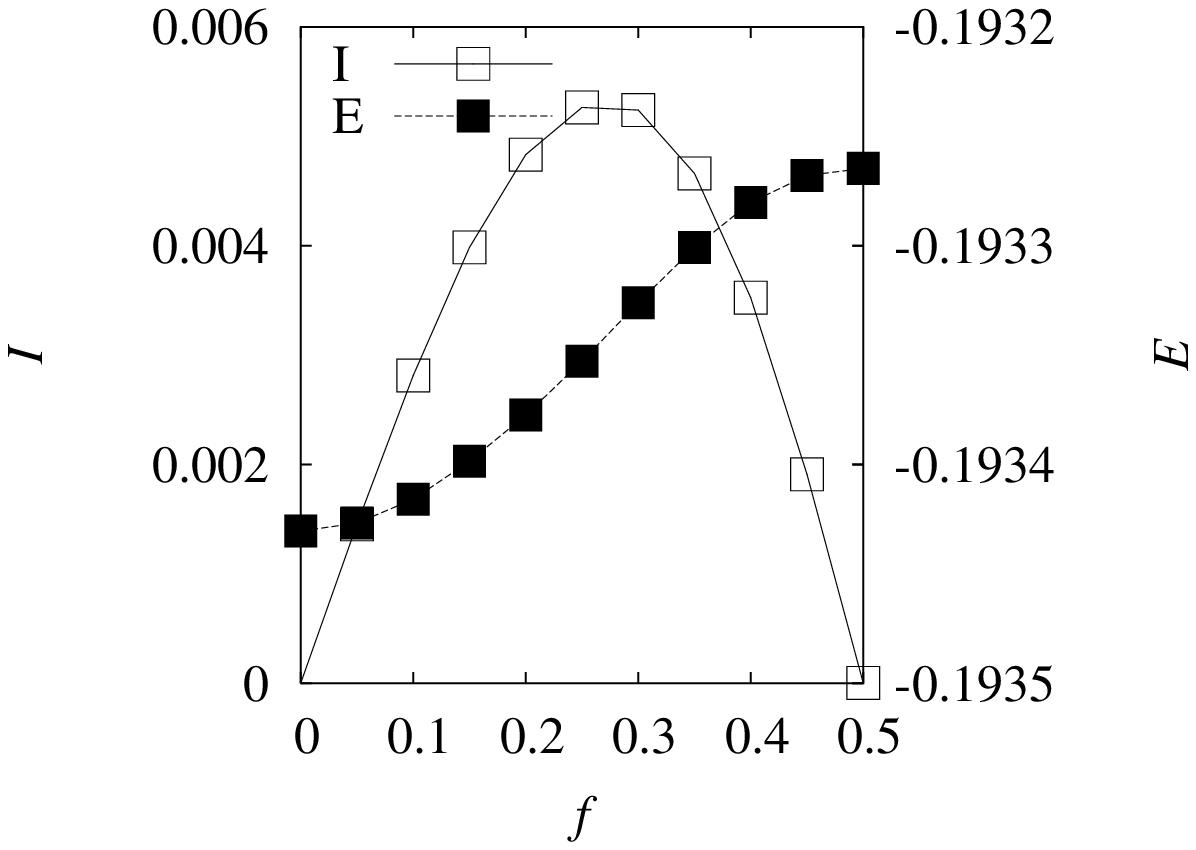,width=6cm}\\ (a) \\
\epsfig{file=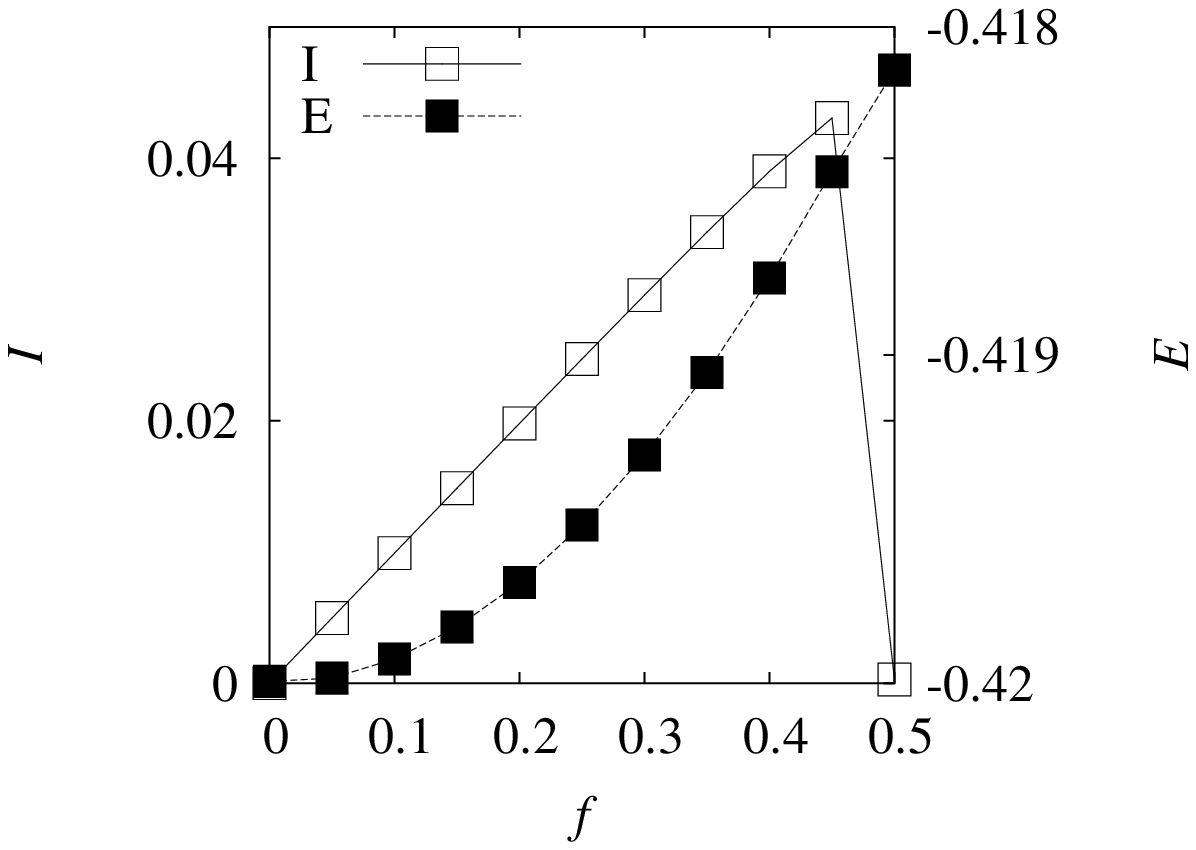,width=6cm}\\ (b)
\caption{Persistent current $I$ and ground state energy $E$ as
  functions of the flux $f$ in (a) the insulating phase ($E_J/8E_0 =
  0.64$) and (b) the superfluid phase ($E_J/8E_0 = 1$) along the
  particle-hole symmetry line ($n_g=0$).}
\label{fig(ccjjn):jjnpcf}
\end{figure}

\begin{figure}[!t]
\centering%
\epsfig{file=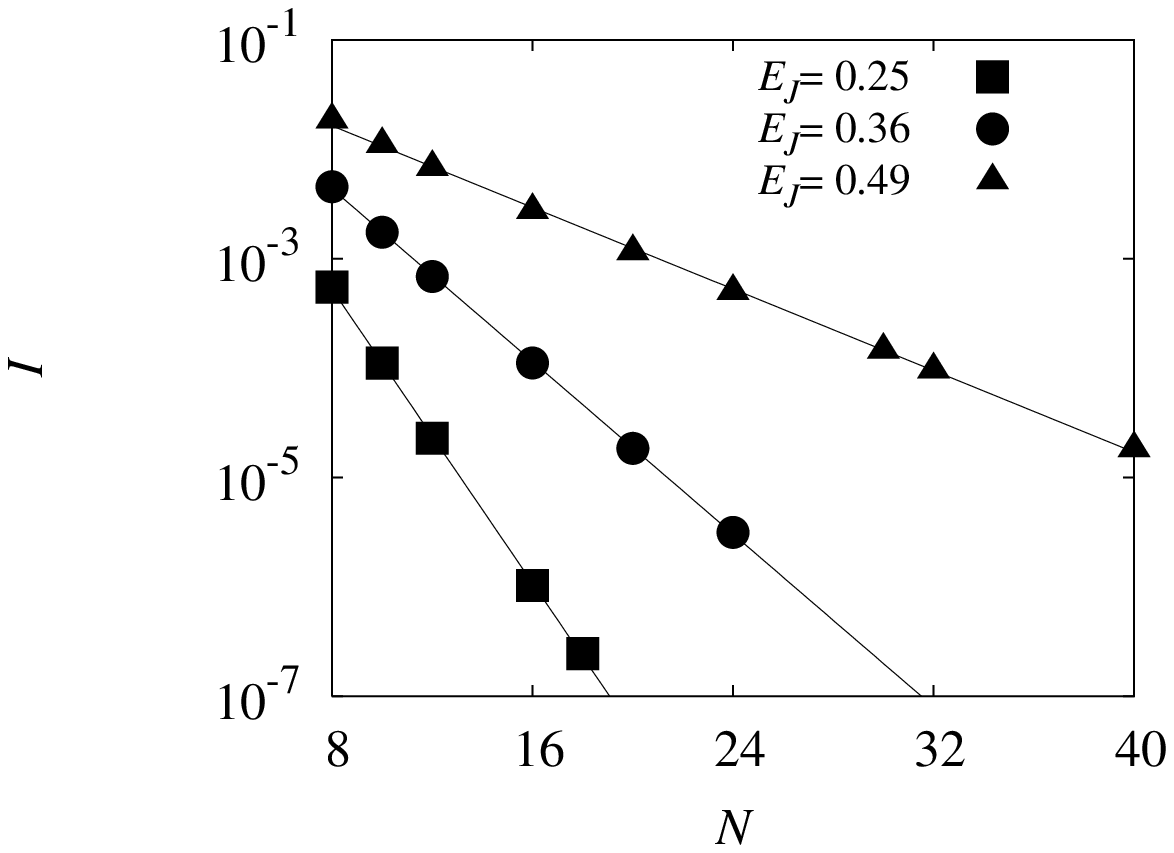,width=6cm}\\ (a) \\
\epsfig{file=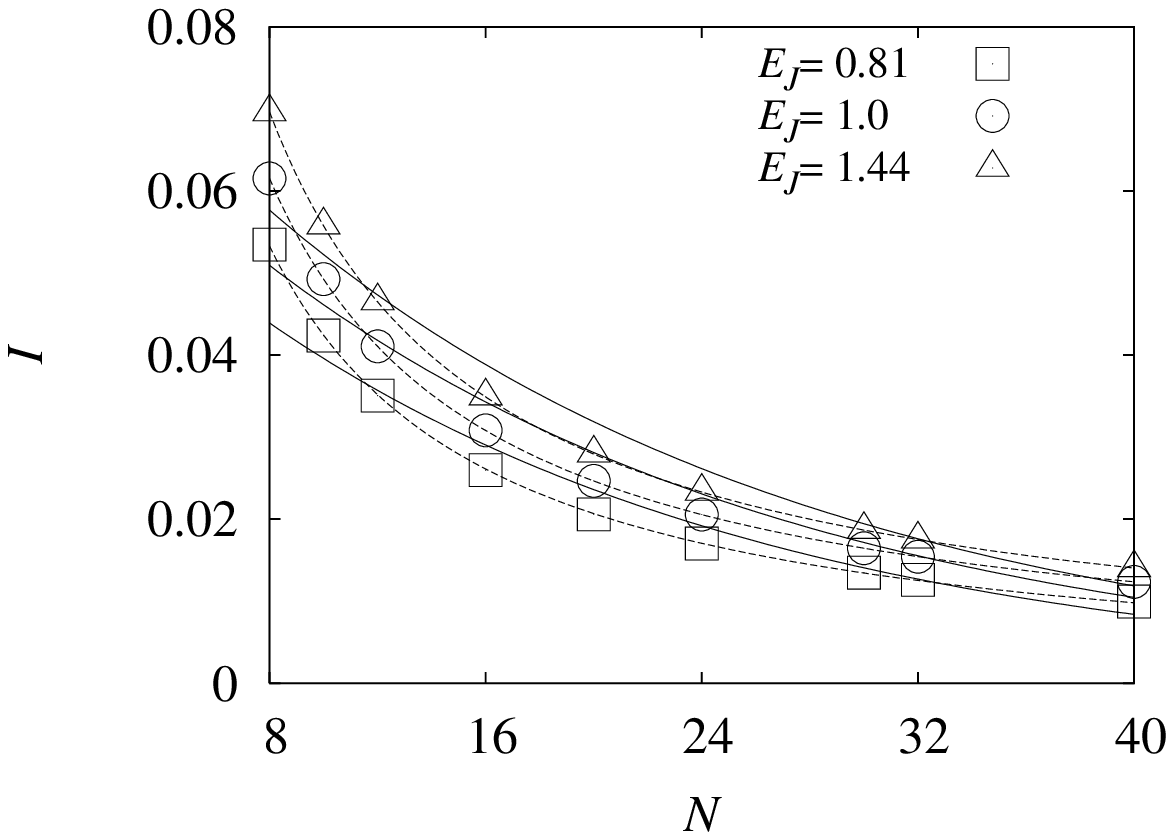,width=6cm}\\ (b)
\caption{Persistent current $I$ versus the system size $N$ in 
  (a) the insulating phase [$E_J = 0.25(\blacksquare),\,
  0.36(\bullet)$, and $0.49(\blacktriangle)$ in units of $8E_0$] and
  (b) the superfluid phase [$E_J = 0.81(\square)$, $1(\circ)$, and
  $1.44(\vartriangle)$ in units of $8E_0$] along the particle-hole
  symmetry line ($n_g=0$).  Solid and dotted lines represent the
  exponential function $a e^{-bN}$ and the algebraic function $c/N$,
  respectively, where the constants $a$, $b$, and $c$ are obtained via
  a fitting algorithm.}
\label{fig(ccjjn):jjnpcN}
\end{figure}

We have calculated the persistent current in a finite-size system
($N=40$) under periodic boundary conditions. The persistent current is
evaluated in the ground state, namely, the lowest-energy state, which
is found by varying the total excess boson number $M$ at given $n_g$
and $E_J$, and expressed in units of $eE_J/N\hbar$ in all subsequent
figures.  \Figsref{fig(ccjjn):jjnpcf}(a) and
\ref{fig(ccjjn):jjnpcf}(b) show the dependence of the persistent
current (in units of $eE_J/N\hbar$) and of the ground state energy (in
units of $8E_0$) on the flux $f$ in both the insulating phase and in
the superfluid phase, respectively, without the gate charge ($n_g=0$).
For small $E_J$, the current depends sinusoidally on $f$, whereas it
has a saw-tooth shape in the superfluid phase. Such behavior of the
persistent current is well known in the two extreme one-dimensional
electron models: In the tight-binding model with the lattice potential
energy dominant over the kinetic energy, the single-particle energy is
given by a cosine function of the flux $f$, giving rise to sinusoidal
dependence of the current on $f$. On the other hand, the free electron
model on a ring, where the energy is quadratic in $f$, has the
persistent current linear in $f$ and of the saw-tooth shape.  In our
model Cooper pairs take the role of the electrons and in analogy we
infer that the saw-tooth dependence in the persistent current
indicates the emergence of the superconductivity over the system,
where the Cooper pairs can freely move around.  Our data for the
dependence of the current on the system size, shown in
\figref{fig(ccjjn):jjnpcN}, also leads to the same interpretation: In
the insulating phase the bosons are localized at sites so that the
probability for a boson to circle around the ring and to return to its
starting position is proportional to $t^N$, where $t$ is the hopping
probability between nearest neighbors.  This gives the current
decaying exponentially with the system size $N$ [see
\figref{fig(ccjjn):jjnpcN}(a)]. On the other hand, in the superfluid
phase the wave function of the boson is extended and the hopping
probability over the system does not depend on the system size.
Instead, since the energy itself is quadratic in the system size, the
persistent current follows a power-law with respect to the system
size, as shown in \figref{fig(ccjjn):jjnpcN}(b). Hence our data for
the persistent current are fully consistent with the phase transition
explained in \secref{sec(ccjjn):qpt_jjn}.

\begin{figure}[!t]
\centering%
\epsfig{file=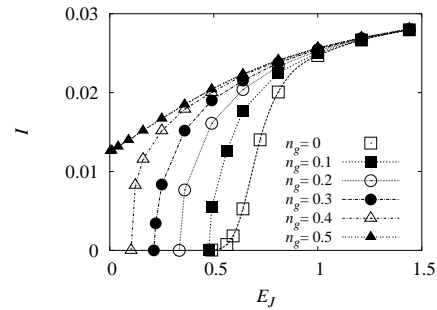,width=6cm}
\caption{Persistent current versus the Josephson energy for various
  gate voltages.  In each case the current, scaled in units of
  $eE_J/N\hbar$, makes clear the contribution of the correlations
  between nearest-neighboring grains.  All the currents are calculated
  at $f=1/4$.}
\label{fig(ccjjn):jjnpc}
\end{figure}

We exhibit the dependence of the persistent current on the Josephson
energy at $f=1/4$ and various gate voltages in
\figref{fig(ccjjn):jjnpc}.  The persistent current scaled by $E_J$ is
negligibly small in the insulating region, then rises rapidly near the
transition point, and increases only marginally in the superfluid
phase.  For $n_g=0$, the current shows finite-size effects, gradually
increasing quite before the transition point. On the contrary, in the
presence of nonzero gate voltage, the current increases very sharply
at the phase boundary even in the small-size system, which is
attributed to the abrupt change in the total boson number or the
density of the ground state at the transition point.  Deep in the
superfluid phase, on the other hand, the persistent current becomes
independent of the gate voltage.

\subsection{Strongly coupled Josephson-junction necklaces}

\begin{figure}[!t]
\centering%
\epsfig{file=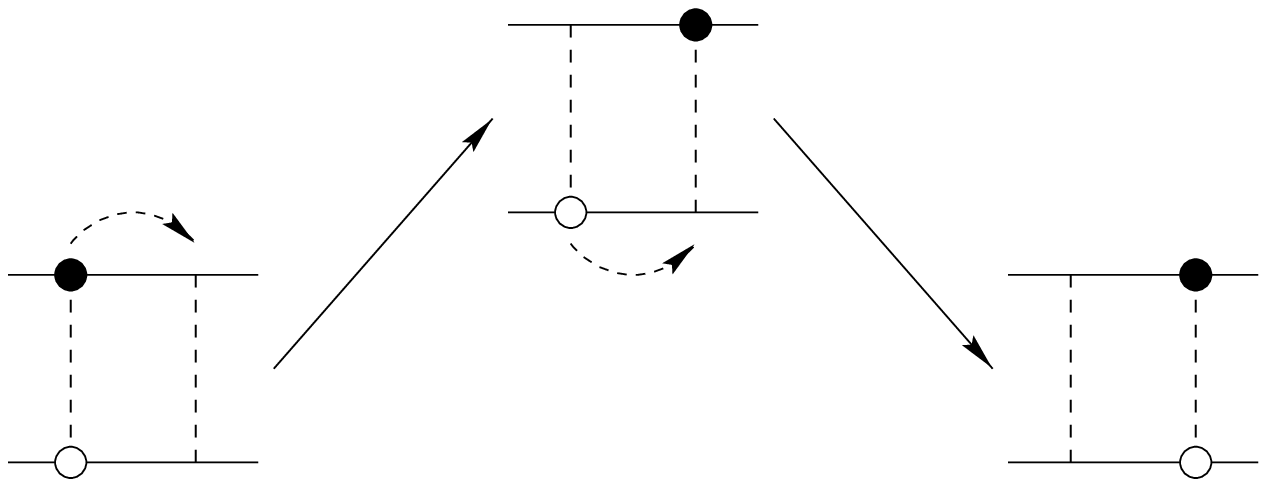,width=6cm}\\ (a) \\[1cm]
\epsfig{file=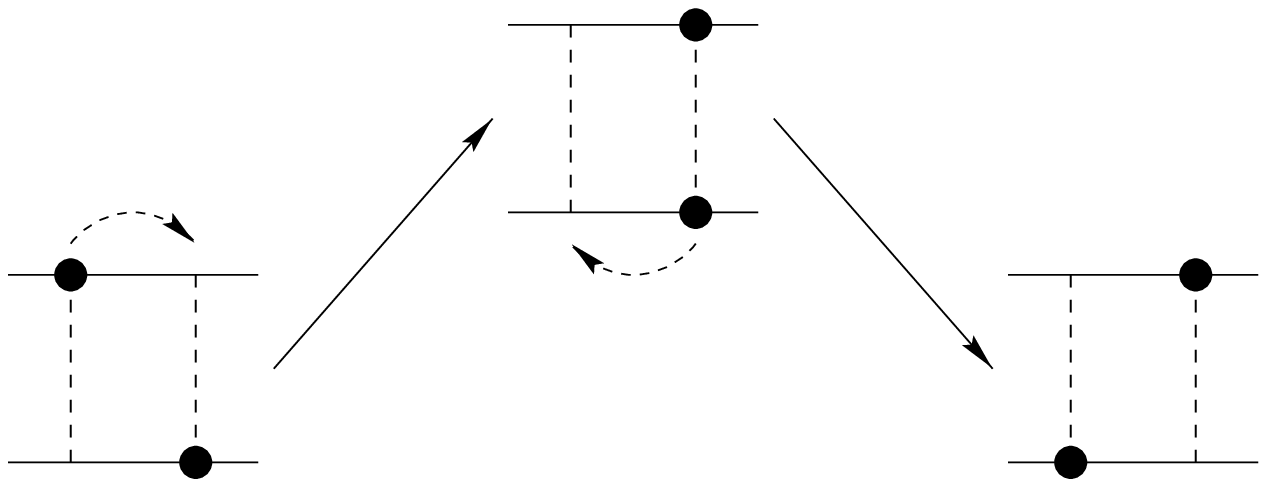,width=6cm}\\ (b)
\caption{Typical cotunneling processes relevant (a) near the
  particle-hole symmetry line and (b) near the maximal-frustration
  line. Such cotunneling results in the current mirror effects.}
\label{fig(ccjjn):cotunnel}
\end{figure}

From the observation in \secref{sec(ccjjn):ptccjjn}, it is evident
that in the strongly coupling limit the excitons such as particle-hole
and particle-void pairs play dominant roles in the transport.  In the
picture of the lowest-order cotunneling processes illustrated in
\figref{fig(ccjjn):cotunnel}, however, such pairs are tightly bound
throughout the transport process.  Accordingly, the current induced in
one necklace is accompanied by the secondary current in the other
necklace, with the same magnitude but in the opposite
direction.\cite{ChoiMS1D98} On the other hand, in response to the
magnetic flux, the charges in an excitonic pair tend to move in
opposite directions since their signs are opposite (with respect to
the offset charge $n_g$).  Therefore the current mirror effect
competes with the influence of the magnetic flux.

\begin{figure}[!t]
\centering%
\epsfig{file=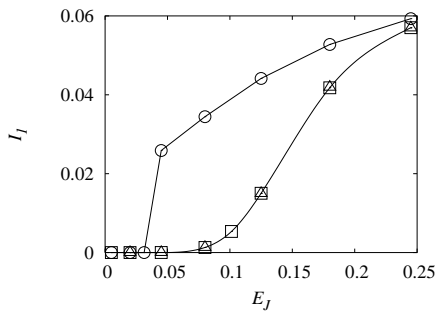,width=6cm}
\caption{Persistent current $I_1$ (along one necklace) versus the
  Josephson energy $E_J$ in the strongly coupled system of size $N=8$,
  for $f=1/4$ and gate voltage $n_g = 0(\square),\, 0.2(\circ)$, and
  $0.5(\triangle)$.  Lines are merely guides to eyes.}
\label{fig(ccjjn):ccjjnpc}
\end{figure}

Indeed, for small values of the Josephson energy ($E_J/8E_0 \lesssim
0.07$), the persistent current is quite negligible both on the
particle-hole symmetry line ($n_g = 0$) and on the maximal-frustration
line ($n_g =1/2$) (see \figref{fig(ccjjn):ccjjnpc}).  The small but
still non-zero amount of persistent current is induced via
higher-order tunneling processes. Namely, the charges in an excitonic
pair break up, run down the circumferences in the opposite directions,
and recombine.  Contributions from these processes are observable only
in a system with a small number of sites ($N=8$ for the data in
\figref{fig(ccjjn):ccjjnpc}).  In other words, the current mirror
effect wins the competition.  It is distinguished from the behavior of
the persistent current in a single Josephson-junction necklace, where
for $n_g = 0$ the current increases rapidly near the transition point
and any strength of the Josephson energy induces rather large
persistent current for $n_g = 0.5$ (see \figref{fig(ccjjn):jjnpc}),
demonstrating the action of a different kind of charge fluctuations in
the coupled system.

For larger values of the Josephson energy ($E_J/8E_0 \gtrsim 0.07$),
however, a considerable amount of persistent current flows through the
system and increases with $E_J$.  It can be explained by the
generation of excitations with higher charging energies in the
presence of the Josephson energy. As observed in
\figref{fig(ccjjn):P}, with the increased Josephson energy, more of
the charge states that do not satisfy $n_x^+ = 0$ (near the
particle-hole symmetry line) or $n_x^+ = 1$ (near the
maximal-frustration line) are now mixed with the lowest charging
energy states.  These excitations can carry a finite amount of
persistent current since the signs of the charges in a pair are not
opposite now.  In short, the magnetic frustration wins the competition
with the current mirror effect.  It is also interesting that the
persistent currents for $n_g = 0\,(\square$) and for
$1/2\,(\triangle$) are almost the same. In fact, the densities of
charge excitations which do not satisfy the lowest-charging energy
condition are nearly the same for the two cases since the charging
energy costs for such excitations amount to the same energy $U_0$ in
both cases.

Note also that unlike these two cases the persistent current for
$n_g=0.2\,(\circ$) increases sharply at the transition point and
becomes quite larger than the one for $n_g = 0$ or $1/2$.
Intermediate values of charge frustration (in the superfluid phase)
bring about a variety of charge excitations in the presence of the
Josephson energy and diminish the energy gap between the charge
excitations, giving rise to the reduction of the current mirror effect
and favoring independent single-charge transport.

\begin{figure}[!t]
\centering%
\epsfig{file=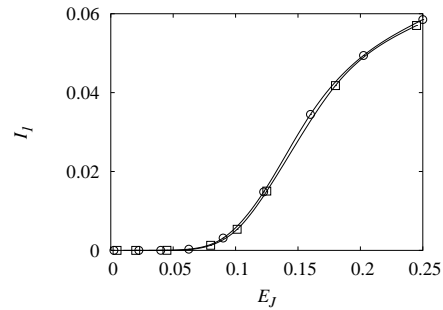,width=6cm}
\caption{Persistent current versus the Josephson energy on the
  particle-hole symmetry line ($n_g = 0$) in the strongly coupled
  system with the coupling capacitance $C_I/C_0 = 100\,(\square)$ and
  $200\,(\circ)$.}
\label{fig(ccjjn):ccjjnpc_CI}
\end{figure}

\Figref{fig(ccjjn):ccjjnpc_CI} shows that the persistent current
increases slightly as the coupling capacitance is raised. On one hand,
a larger value of the coupling capacitance reduces the lowest
excitation energy $(\sim E_I)$ and makes the excitons more proliferate
in the system, thus increasing the persistent current due to the
excitons. On the other hand, breaking of the excitons, which is
crucial for inducing the persistent current, costs higher energy
$(\sim E_0)$.  These conflicting trends result in the slight increase
in the regime of our interest $(E_0\gg E_I)$.

\begin{figure}[!t]
\centering%
\epsfig{file=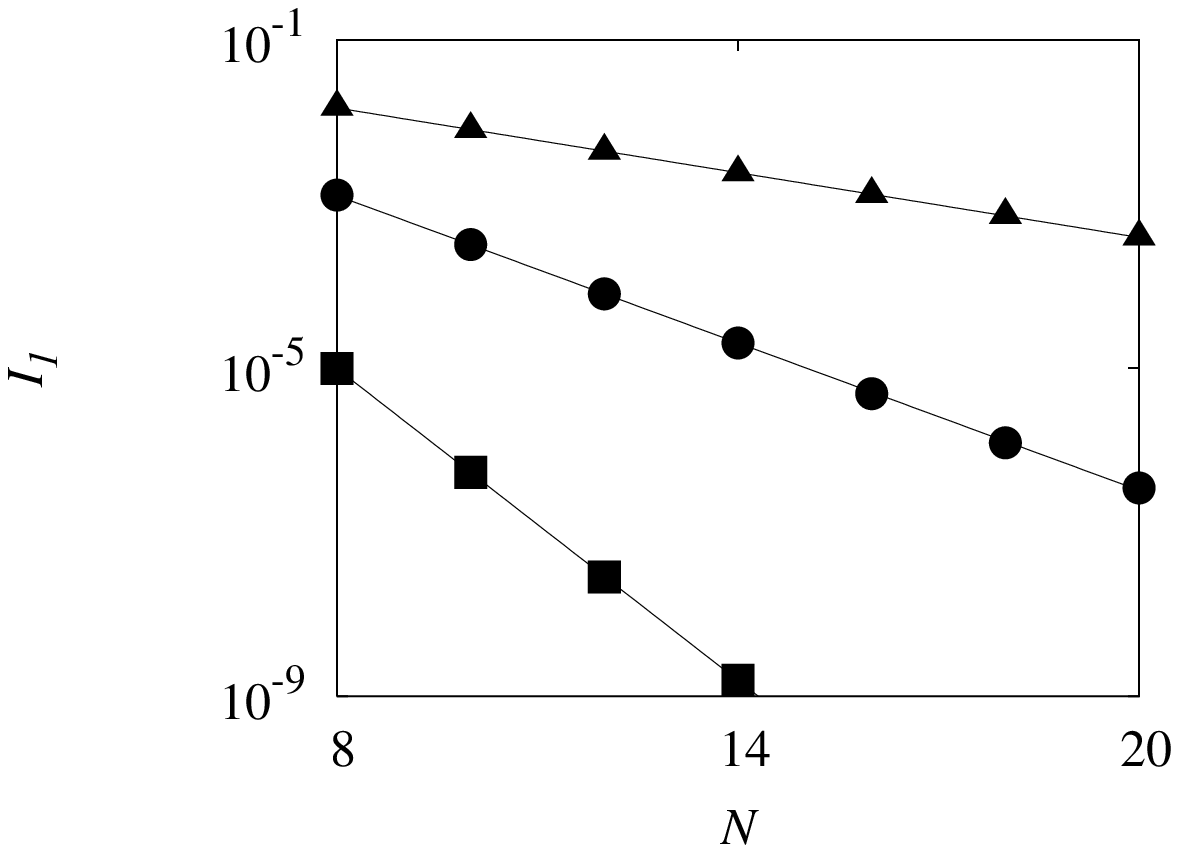,width=6cm}\\ (a) \\
\epsfig{file=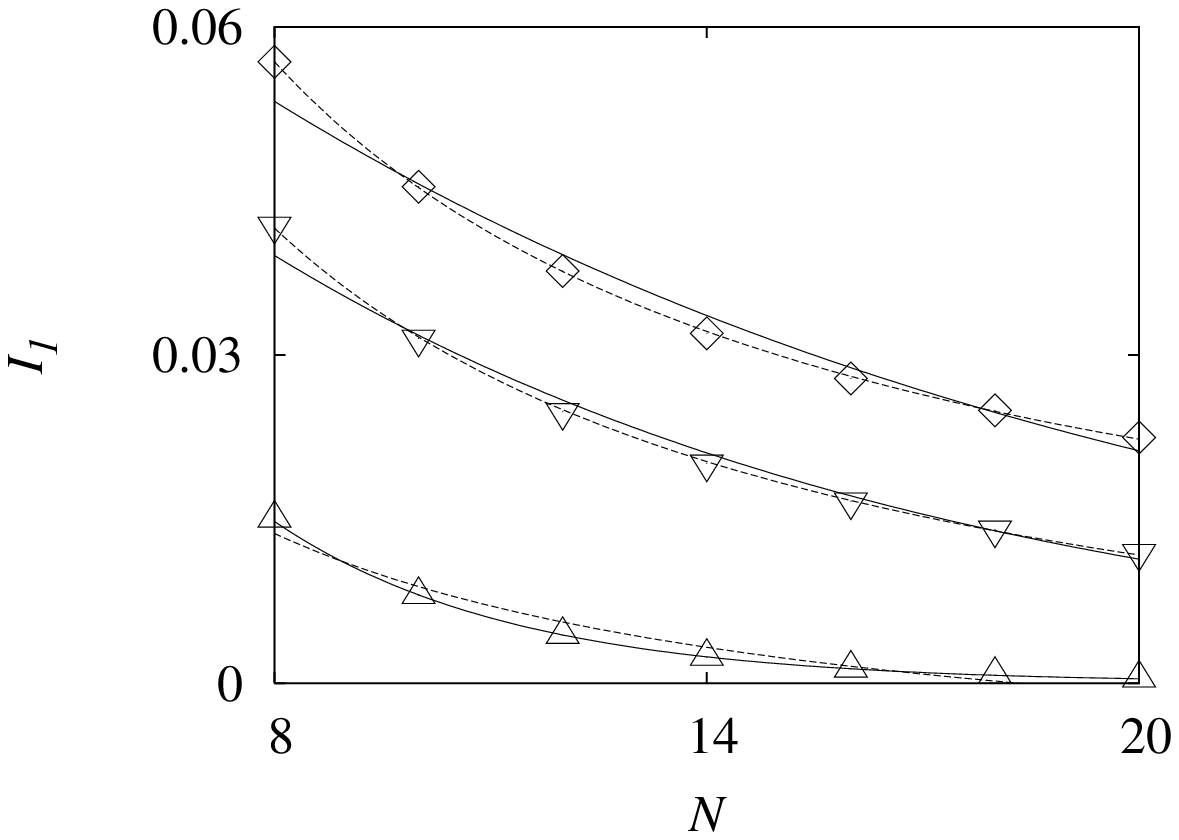,width=6cm}\\ (b)
\caption{Persistent current versus the system size on the particle-hole
  symmetry line ($n_g=0$) for different values of the Josephson energy
  $E_J =$ (a) $0.045\,(\blacktriangle)$, $0.08\,(\bullet)$, and
  $0.125\,(\blacktriangle)$; (b) $0.125\,(\vartriangle)$,
  $0.18\,(\triangledown)$, and $0.245\,(\lozenge)$ (again in units of
  $8E_0$). The fitting curves are the same as those given in
  \figref{fig(ccjjn):jjnpcN}.  }
\label{fig(ccjjn):ccjjnpcN}
\end{figure}

The dependence of the current on the system size also supports our
scheme for the role of the excitons in the persistent current.
\Figref{fig(ccjjn):ccjjnpcN} shows that, similarly to the case of a
single necklace, in the insulating phase the current decays
exponentially with the system size and decreases inversely to the
system size deep in the superfluid phase. Near the transition point on
the side of the superfluid phase (at $E_J/8E_0 = 0.08$ and $0.125$),
however, the current does display exponential dependence on the system
size, which indicates that spatially localized objects participate in
the generation of the current. It is another evidence for the virtual
processes of unpaired charges or higher-order excitons, in the region
where the low-lying excitons themselves are delocalized over the
system.

\begin{figure}[!t]
\centering%
\epsfig{file=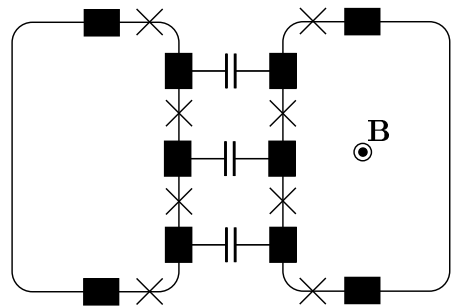,width=6cm}
\caption{Schematic diagram of two coupled Josephson-junction
  necklaces.  Part of the two necklaces are capacitively connected
  whereas the magnetic field threads only one of the two necklaces.}
\label{fig(ccjjn):necklaces_asym}
\end{figure}

To reveal the cotunneling process more explicitly, we devise another
interesting configuration that the magnetic field penetrates only one
necklace ($l=1$) without affecting the other ($l=2$).  Such a setup
may be realized experimentally as shown in
\figref{fig(ccjjn):necklaces_asym}.  Notice that only part of the two
necklaces are capacitively connected.  In order for the persistent
current to flow through uncoupled grains, the Josephson coupling
between those uncoupled grains should be sufficiently large.  In this
arrangement, one can observe the current mirror effect, similar to the
case that only one chain is biased by an external
voltage.\cite{Matters97,Shimada00}

\begin{figure}[!t]
\centering%
\epsfig{file=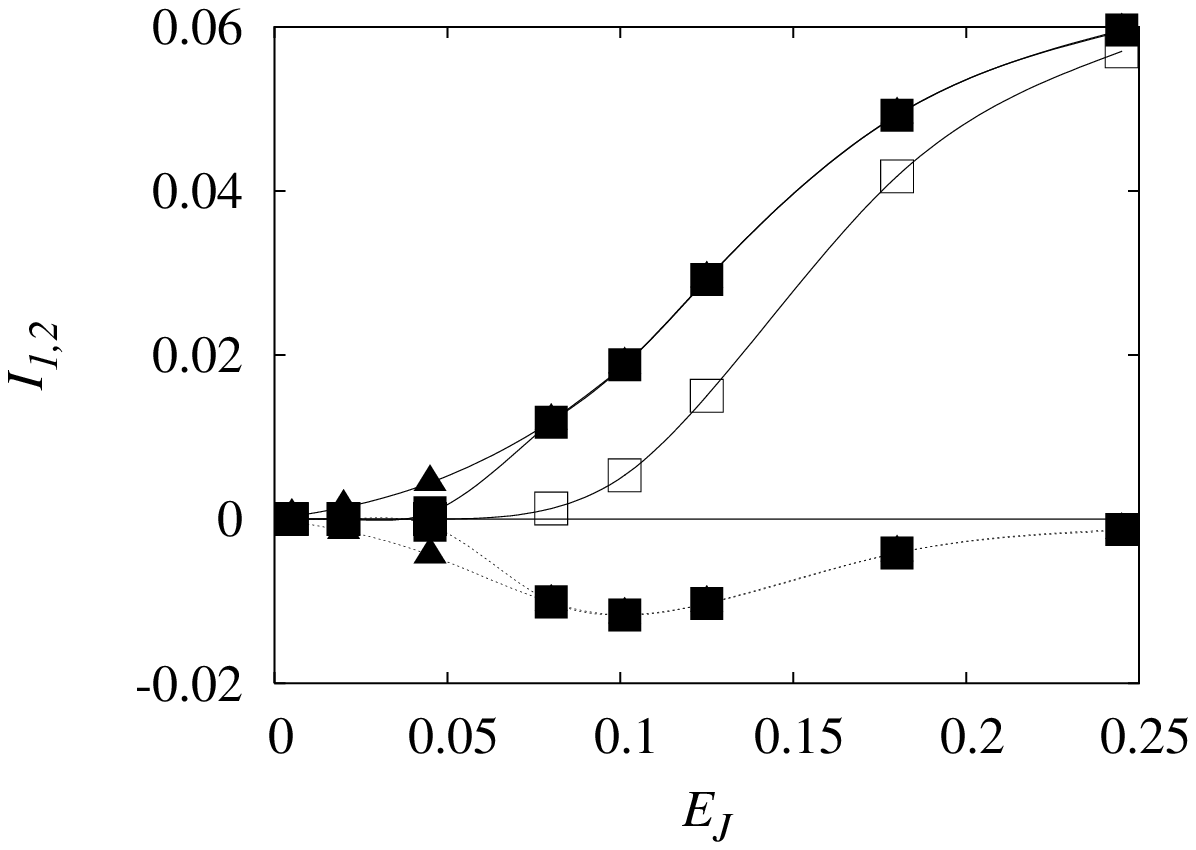,width=6cm}\\ (a) \\
\epsfig{file=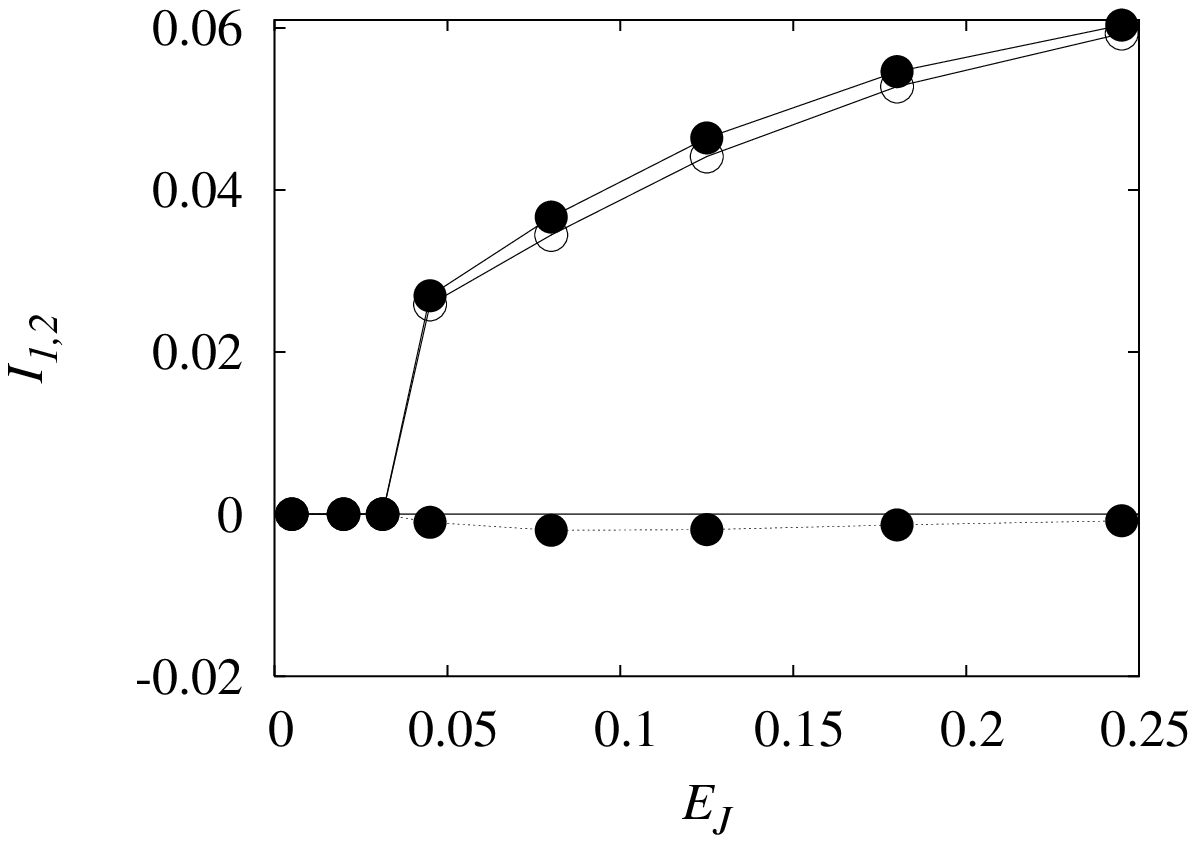,width=6cm}\\ (b)
\caption{Persistent currents $I_1$ (solid lines with filled symbols)
  and $I_2$ (dotted lines with filled symbols) when the magnetic field
  threads only the first $(l=1)$ necklace.  The currents are
  calculated for the gate voltage $n_g =$ (a) $0\,(\blacksquare),\,
  0.5\,(\blacktriangle)$; (b) $0.2\,(\bullet)$.  For comparison, the
  persistent currents when both the necklaces are threaded by the
  magnetic field (see \figref{fig(ccjjn):ccjjnpc}) are also plotted,
  represented by the corresponding empty symbols.}
\label{fig(ccjjn):ccjjnpc_asym}
\end{figure}

\Figref{fig(ccjjn):ccjjnpc_asym}(a) exhibits the current mirror effect
in the system with the magnetic field threading only the first $(l=1)$
necklace.  On the particle-hole symmetry and the maximal-frustration
lines, the two persistent currents $I_1$ and $I_2$ along the first and
the second $(l=2)$ necklaces, respectively, satisfy the relation $I_1
\approx -I_2$ in the range $E_J/8E_0 \lesssim 0.07$. As $E_J$ is
increased further, nonetheless, not only the mirror effect disappears
gradually but also the current $I_2$ diminishes to zero. It means that
the independent single-charge transport rather than the cotunneling
transport is favorable at large values of the Josephson energy.  Note
that the current $I_1$ is much higher than the corresponding current
in the system with the magnetic field acting on both necklaces.
Interestingly, unlike the previous setup, the persistent current for
$n_g = 1/2$ is higher than that for $n_g = 0$.  Whereas on the
maximal-frustration line the system is in the superfluid state of the
particle-void pairs even at small $E_J$, on the particle-hole symmetry
line a sufficient amount of the Josephson energy is necessary for
generating excitons, i.e., particle-hole pairs.  For $n_g = 0.2$, the
current mirror effect is indeed negligible and the increase in $I_1$
is also very small, as shown in \figref{fig(ccjjn):ccjjnpc_asym}(b).

\section{Conclusion\label{sec(ccjjn):conclusion}}

We have studied phase transitions and persistent currents in a ladder
of two capacitively coupled Josephson-junction necklaces.  Emphasis
has been paid on the roles of excitons in the presence of charge and
magnetic frustration.  To obtain the properties of the ground and
excited states of the system, we have utilized the DMRG method for
arbitrary values of the gate charge and of the magnetic flux.
Although the main interest lies in the strong-coupling limit between
the two necklaces, we have studied both the uncoupled and strongly
coupled cases for comparison.  In both cases, the gate voltage brings
about crucial effects on the properties of the system.  In a single
Josephson-junction necklace, the presence of the gate voltage changes
rather abruptly the behavior of the persistent current as well as
nature of the phase transition.  On the other hand, in the
capacitively coupled Josephson-junction necklaces, such drastic change
is not observed but the gate voltage determines the class of excitons
driving the phase transition: the particle-hole pairs near the
particle-hole symmetry line and the particle-void pairs near the
maximal-frustration line. In the presence of the Josephson tunneling,
two different superfluid phases, characterized by the condensation of
either of the two types of excitons, have been identified, depending
on the gate charge. The pair correlation function and the exciton
density have provided evidence for the formation of such excitons.

In the strongly coupled necklaces, the behavior of the persistent
current is manifested by the competition between the current mirror
effect and magnetic frustration, which is associated with the
cotunneling transport of the bound excitonic pairs of either particles
and holes or particles and voids.  At small values of the Josephson
energy, the current mirror effect wins the competition and only a very
small amount of persistent current is allowed for a finite-sized
system.  At large values of the Josephson energy, magnetic frustration
can make use of higher-charging-energy states to dominate over the
current mirror effect, allowing a considerable amount of persistent
current.  We have also suggested an experimentally realizable system
to demonstrate the cotunneling process of the excitons.
To out present knowledge, the only experimental work related to the
system considered here is that reported in
Ref.~\onlinecite{Shimada00}.  Unfortunately, however, the large bias
voltage applied to both arrays does not allow us to make a direct
connection.  In particular our DMRG algorithm is not suitable for such
a non-equilibrium problem.

\acknowledgments

We acknowledge the partial supports from the SKORE-A Program and from
the BK21 Program.


\begin{thebibliography}{99}
\bibitem{endnote1} In literature, ``particles'' and ``holes'' are used
  mostly to refer to elementary excitations in fermion systems.  In
  this paper concerning about arrays of superconducting grains, we use
  them to refer to excess and deficit (boson-like) Cooper pairs rather
  than the constituent electrons.
\end{thebibliography}
\end{document}